\documentclass{JFM-FLM_Au}

\makeatletter
\let\oddabsfooterflag\relax
\let\evenabsfooterflag\relax
\let\footerflagdefns\@gobble
\let\oddfooterflagdefns\@gobble
\let\evenfooterflagdefns\@gobble
\makeatother

\usepackage{etoolbox}
\makeatletter
\gdef\@journal{}
\gdef\@underjournal{}
\def\ps@titlepage{\leftskip\z@\let\@mkboth\@gobbletwo\vfuzz=5\p@
  \def\@oddhead{}\def\@evenhead{}%
  \def\@oddfoot{}\def\@evenfoot{}%
  \def\sectionmark##1{}%
  \def\subsectionmark##1{}%
}
\patchcmd{\@maketitle}{(Received xx; revised xx; accepted xx)\hfill}{}{}
  {\@latex@warning{arXiv patch: editorial dates not removed from title page}}
\def\ps@headings{\let\@mkboth\markboth
  \def\@oddhead{\hfill{\itshape\@righttitle}\hfill}%
  \def\@evenhead{\hfill{\itshape\@lefttitle}\hfill}%
  \def\@oddfoot{\hfil\normalfont\small\thepage\hfil}%
  \def\@evenfoot{\hfil\normalfont\small\thepage\hfil}%
  \def\sectionmark##1{\markboth{##1}{}}%
  \def\subsectionmark##1{\markright{##1}}%
}
\makeatother
\usepackage{amssymb}
\usepackage{amsmath}
\usepackage{graphicx}
\usepackage{subfig}
\usepackage{url}
\usepackage{amsfonts}
\usepackage{tikz}
\usepackage{booktabs}
\usepackage{acronym}
\usetikzlibrary{calc}  
\usetikzlibrary{arrows.meta}
\usepackage{xcolor}
\usepackage[normalem]{ulem}

\acrodef{AFC}[AFC]{Active Flow Control}
\acrodef{AE}[AE]{Autoencoder}
\acrodef{CAE}[CAE]{Convolutional Autoencoder}
\acrodef{CNN}[CNN]{Convolutional Neural Network}
\acrodef{DNS}[DNS]{Direct Numerical Simulation}
\acrodef{GTS}[GTS]{Ground Transportation System}
\acrodef{LSTM}[LSTM]{Long Short-Term Memory}
\acrodef{MIMO}[MIMO]{multi-input multi-output}
\acrodef{MLP}[MLP]{multi-layer perceptron}
\acrodef{MPC}[MPC]{Model Predictive Control}
\acrodef{POD}[POD]{Proper Orthogonal Decomposition}
\acrodef{ROM}[ROM]{Reduced-Order Model}
\acrodef{VAE}[VAE]{Variational Autoencoder}
\acrodef{bVAE}[$\beta$-VAE]{$\beta$-Variational Autoencoder}
\acrodef{dklVAE}[DKL-VAE]{Decomposed-KL Variational Autoencoder}
\acrodef{SINDy}[SINDy]{Sparse Identification of Nonlinear Dynamics}
\acrodef{PSD}[PSD]{Power Spectral Density}
\acrodef{RNN}[RNN]{Recurrent Neural Network}

\title{The balance between compactness and forecast accuracy of data-driven latent-space reduced-order models in controlled wake flows}

\lefttitle{A. Solera-Rico, P. Garc\'ia-Caspue\~nas, C. Sanmiguel Vila and S. Discetti}
\righttitle{Compactness versus forecast accuracy in controlled wake-flow ROMs}

\author{Alberto Solera-Rico\aff{1,2},
  Patricia Garc\'ia-Caspue\~nas\aff{3},
  Carlos Sanmiguel Vila\aff{1,2}
 \and Stefano Discetti\aff{2}}

\affiliation{\aff{1}Sub-directorate general of aeronautical systems, Spanish National Institute for Aerospace Technology (INTA), ctra. M-301, km. 10.5, 28330 San Mart\'in de la Vega, Madrid, Spain. ROR: https://ror.org/02m44ak47
\aff{2}Departamento de Ingenier\'ia Aeroespacial, Universidad Carlos III de Madrid, Avenida de la Universidad 30, 28911 Legan\'es, Madrid, Spain. ROR: https://ror.org/03ths8210
\aff{3}Department of Mechanical Engineering, University of Washington, 1410 NE Campus Parkway, Seattle, WA 98195, USA. ROR: https://ror.org/00cvxb145}

\corresau{Stefano Discetti, \email{sdiscett@ing.uc3m.es};
Alberto Solera-Rico,
\email{alberto.solera@alumnos.uc3m.es}}

\begin{document}
\maketitle

\begin{abstract}
Model-based active flow control requires predictive models that are accurate, stable, and fast enough for real-time optimisation. In controlled wake flows, this is often achieved through \acp{ROM} that first compress high-dimensional velocity snapshots into a latent space and then learn a time-stepping predictor for the dynamics in the latent space. Here, we study how the choice of the spatial encoder affects the predictability of the resulting latent coordinates for wake flows under control inputs. Using two actuated 2D wake configurations, a simplified truck wake and the fluidic pinball, we compare \ac{POD} against nonlinear \acp{CAE} and two types of variational autoencoders for compression, and evaluate several temporal predictors based on \acl{LSTM} networks. \acp{CAE} achieve higher compression efficiency and sharper short-term reconstructions, but they produce latent dynamics that are more irregular and with broadband spectral content. As a consequence, long-horizon forecasts degrade faster and show a higher probability of catastrophic divergence than POD-based models. \ac{POD} yields smoother latent trajectories that are easier to learn and extrapolate, leading to more reliable predictions beyond the short-term regime. These results reveal a clear trade-off between compactness and forecast accuracy, and suggest that the stability of the latent dynamics prediction can outweigh maximal compression. This is particularly relevant for control strategies rooted in forecasts of the dynamics, such as model predictive control and reinforcement learning. The findings provide practical guidance for designing actuation-aware, hardware-feasible predictive \acp{ROM} for real-time flow control.

\end{abstract}


\section{\label{sec:introduction}Introduction}

\ac{AFC} has emerged as a relevant strategy to enhance the aerodynamic performance of ground and air vehicles \citep{brunton2015closed}. By manipulating flow separation and wake structures using actuators such as synthetic jets, plasma devices, or moving surfaces, significant performance gains can be achieved~\citep{Greenblatt2000separationcontrol, Glezer2002reviewJets, Cattafesta2011reviewActuators}. For ground vehicles, the efficacy of open-loop actuation has been well-demonstrated in wind tunnel studies on simplified bluff bodies, where steady and pulsed blowing can substantially recover base pressure and reduce drag~\citep{littlewood2012squareback, mcnally2015drag, cerutti2020van, amico2022deep, amico2024flow, robledo2025van}.

Feedback control promises further significant improvements~\citep{brunton2015closed}. Real-world conditions are inherently unsteady, and an effective \ac{AFC} system must be able to sense the current flow state, predict its evolution, and apply an optimal control action that is adapted in real-time. To meet these demands, data-driven methods and machine learning are now opening new research avenues~\citep{Brunton2020mlFluids}. Broadly, these approaches fall into two categories: model-free controllers that learn a policy directly from data, and model-based controllers that exploit an explicit dynamical model.

Model-free strategies can be attractive when little is known \textit{a priori} about the system. Such approaches have shown promising results in low-Reynolds-number or numerically simulated flows where large amount of training data are generated in a cheap and safe manner~\citep{garcia2025deep,suarez2025flow}. However, model-free methods are often sample inefficient, provide less transparent guarantees on stability and safety, and struggle to incorporate hard constraints on actuators or states. These limitations are particularly restrictive in experimental \ac{AFC}, where data is expensive and hardware must operate within strict bounds. This motivates a focus on model-based approaches, where control decisions are derived from an explicit predictive model.

A typical example of model-based control is \ac{MPC}~\citep{rawlings2017book, CamachoBordons2013_MPC}. At each time step, \ac{MPC} leverages a model of the dynamics to forecast future flow states over a finite horizon and solves an optimisation problem to obtain the optimal control sequence. Compared with purely model-free strategies, this framework is more sample efficient, offers clearer routes to stability and safety guarantees, and handles constraints naturally. The primary challenge is therefore the development of a predictive model that is both sufficiently accurate to capture the controlled dynamics and computationally efficient for real-time execution~\citep{kaiser2018sindyMPC,bieker2020deep,Marra2024Self,solera2025framework,liu2025model}.

The effectiveness of \ac{MPC} ultimately hinges on the reliability of the plant model. In turbulent flows, the difficulty is not only to approximate the unforced dynamics, but also to represent how actuation reshapes the accessible state space. Recent studies show that controlled wakes often evolve on low-dimensional, actuation-dependent manifolds~\citep{marra2024actuation}. Reduced coordinates learned without accounting for control inputs may thus yield latent dynamics that are unnecessarily complex or even inconsistent with feasible actuation commands. This perspective motivates the need to develop actuation-aware reduced-order predictors that remain simple enough for receding-horizon optimisation while capturing the key controlled dynamics.

For this reason, this work focuses on the central challenge of building predictive models suitable for model-based control. While being a gold standard and highly informative in predictive-control implementations ~\citep{bewley2001dns}, full-order simulations are computationally intractable in real-time loops. We thus focus on a framework based on \acp{ROM}, with two stages: (i) compress the high-dimensional velocity field into a low-dimensional latent space (the \textit{encoding} stage), and (ii) learn a time-stepping model that predicts the latent evolution under actuation (the \textit{prediction} step). This approach has been recently explored by several studies with data-driven methods~\citep{maulik2021reduced,bukka2021assessment,solera2024bvae}. This separation is consistent with many classical \ac{ROM} pipelines, where the compression step is based on \ac{POD}: since \ac{POD} is a purely energy-based decomposition, the resulting latent basis is not explicitly optimised jointly with the subsequent dynamical estimator.  While the merits of such techniques are already apparent, we identify two research gaps. On the one hand, the generalisation capabilities of these methods under different exogenous inputs are not clear, and a recipe for training and predicting under environments with active control has not been consolidated yet. On the other hand, there is a growing tendency towards using powerful encoders, while it is not clear if an Occam's razor solution could be more effective under some configurations. Wake flows, for instance, often evolve on low-dimensional attractors, with a rather predictable dynamics. In flows dominated by vortex shedding, the leading \ac{POD} coefficients often exhibit narrow-band, nearly oscillatory dynamics because \ac{POD} isolates the most energetic coherent structures. In special cases, such as nearly periodic flows, these structures may align with Koopman-related spectral components \citep{rowley2009spectral,mezic2013review,taira2017modalanalysis}. However, it must be remarked that, even though \ac{POD} is a linear decomposition, it does not imply that a linearised dynamics is achievable in its reduced space. Therefore, any analogy between the dynamics of the leading \ac{POD} coefficients and Koopman modes should be understood as problem-dependent rather than intrinsic to \ac{POD}. On the other hand, \acp{CAE} can be driven to approximate Koopman operators \citep{Lusch2018DeepKoopman}, although, generally speaking, training driven by reconstruction compactness does not entail robustness guarantees on the dynamics. Existing comparisons between \ac{POD} and \ac{CAE} have been targeted already for systems with autonomous dynamics \citep{Fresca2021DLROM,Fresca2022PODDLROM}; however, the interaction between the encoder choice and multi-input actuation, on actuation-dependent manifolds, remains less explored.

To address the gaps above, we compare linear and nonlinear options for encoding, considering wake flow systems under control actions. For encoding, we contrast the classical energy-optimal \ac{POD}~\citep{sirovich1987turbulence} with nonlinear \acp{CAE}~\citep{solera2024bvae}, which typically yield more compact latent representations but may induce more complex controlled latent dynamics. For temporal prediction, we employ \ac{LSTM} networks~\citep{Hochreiter1997lstm}, which are well suited for sequential latent dynamics. Using two 2D wake flows, a simplified truck model~\citep{solera2025framework} and the fluidic pinball~\citep{deng2020}, we quantify the trade-off between compactness and forecast accuracy of the latent space dynamics, and show how the preferred reduction strategy depends on the prediction horizon required by the controller.

The paper is organised as follows. Section \ref{sec:methodology} provides a description of the methodology, including the generation of data for the chosen test cases. The analysis of the model results is provided in \S~\ref{sec:results}. Finally, the conclusions are discussed in \S~\ref{sec:conclusion}.

\section{\label{sec:methodology}Methodology}

\begin{figure}[!ht]
    \centering
    \includegraphics[width=\linewidth]{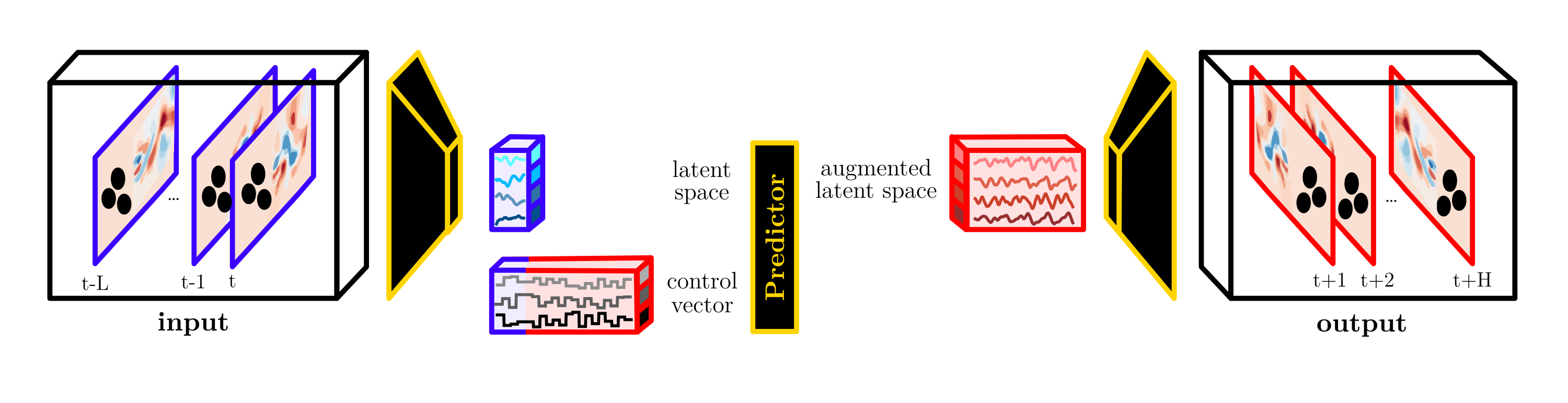}
    \caption{Schematic of the data-driven reduced-order modelling framework. High-dimensional input velocity fields are compressed into a low-dimensional latent-space time-series using an encoder (\ac{POD} or \ac{CAE}). This latent state, along with the corresponding control sequence, is fed into a temporal predictor to infer the future evolution of the latent space. Finally, a decoder reconstructs the predicted latent states back into the high-dimensional, full-field velocity state space.}    \label{fig:2_methodology_scheme}
\end{figure}

This section details the methodology to develop and evaluate predictive reduced-order models. First, we describe the two flow configurations and the generation of their respective high-fidelity \ac{DNS} datasets. Next, we present the two methods used to encode the high-dimensional spatial fields into a low-dimensional latent-space. Finally, in \S~\ref{subsec:predictors}, we detail three approaches based on \ac{LSTM} network architectures for predicting the temporal evolution of this latent space in response to control inputs. The proposed \ac{ROM} framework is illustrated schematically in Figure~\ref{fig:2_methodology_scheme}.

\subsection{Flow configurations and data generation}

Two different 2D wake flow configurations are used to assess the control architectures. These cases are chosen to represent different control challenges: the first is a single-input system, while the second is a more complex multi-input problem. Both datasets are created using \acp{DNS} with open-loop control signals designed to excite the underlying flow dynamics. In low-Reynolds-number flows, \ac{DNS} provides a simulation framework with minimal numerical approximations to precisely represent flow dynamics and capture all scale features. The following sections describe the setup for each case in detail, a summary of the datasets can be found in table~\ref{tab:dataset_summary}.

\subsubsection{Simplified 2D truck model}

\begin{figure}[ht!]
\centering
\begin{tikzpicture}[
    scale=1.3, 
    annot/.style={-{Stealth[length=2mm, width=1.5mm]}},
    every node/.style={font=\sffamily\footnotesize} 
]

\def\rectwidth{7.65}
\def\rectheight{1}
\def\radius{0.118}

\def\jetwidth{0.05}
\def\jetlength{0.3} 

\draw[thick] 
    (\rectwidth, \rectheight/2) 
    -- (\radius, \rectheight/2) 
    arc(90:180:\radius)
    -- (0, -\rectheight/2 + \radius)
    arc(180:270:\radius)
    -- (\rectwidth, -\rectheight/2)
    -- cycle;

\fill[blue, opacity=0.8]
    (\rectwidth, \rectheight/2) 
    .. controls (\rectwidth + \jetlength, \rectheight/2) and (\rectwidth + \jetlength, \rectheight/2 - \jetwidth) .. (\rectwidth, \rectheight/2 - \jetwidth)
    -- cycle;

\fill[blue, opacity=0.8]
    (\rectwidth, -\rectheight/2)
    .. controls (\rectwidth - \jetlength, -\rectheight/2) and (\rectwidth - \jetlength, -\rectheight/2 + \jetwidth) .. (\rectwidth, -\rectheight/2 + \jetwidth)
    -- cycle;
    
\draw (0,-\rectheight/2) -- ++(0,-0.5); 
\draw (\rectwidth,-\rectheight/2) -- ++(0,-0.5); 
\draw[{Latex[length=1.5mm]}-{Latex[length=1.5mm]}] (0,-0.8) -- node[below] {$L=7.65W$} (\rectwidth,-0.8);

\draw (0, \rectheight/2) -- ++(-0.5, 0); 
\draw (0, -\rectheight/2) -- ++(-0.5, 0); 
\draw[{Latex[length=1.5mm]}-{Latex[length=1.5mm]}] 
    (-0.3, -\rectheight/2) -- node[left] {$W$} (-0.3, \rectheight/2);

\coordinate (R_annot_start) at (-\rectheight*.25, \rectheight/2 + 0.4);
\draw[annot] (R_annot_start) -- (0, \rectheight/2);
\node[left=2pt] at (R_annot_start) {$r=0.118W$};

\coordinate (Wj_annot_start) at (\rectwidth-\rectheight*.25, \rectheight/2 + 0.4);
\draw[annot] (Wj_annot_start) -- (\rectwidth, \rectheight/2);
\node[left=1pt] at (Wj_annot_start) {$w_{jet}= 0.05W$};

\draw[-{Stealth[length=4mm, width=3mm]}, line width=1.25pt] (-1.8*\rectheight, 0) -- (-\rectheight, 0);
\node[above=2pt] at (-\rectheight*1.5, 0) {$U_{\infty}$};

\end{tikzpicture}
\caption{Schematic of the simplified truck flow configuration. Zero-net-mass flow jets are illustrated in blue.}
\label{fig:TruckSchematic}
\end{figure}
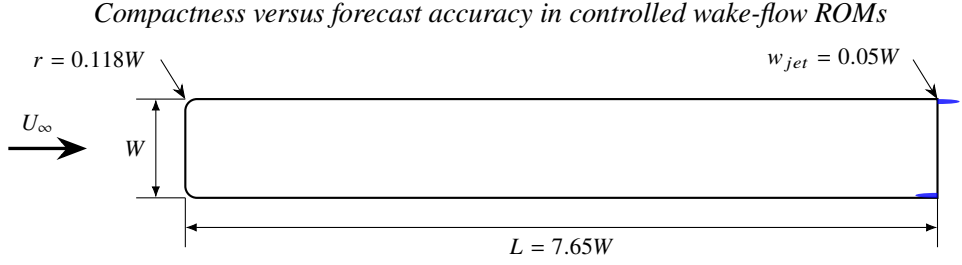

This first case is a simplified model of a wake flow around a road vehicle. Figure~\ref{fig:TruckSchematic} depicts the flow configuration based on the horizontal mid-plane geometry of the \ac{GTS} \citep{GroundTransportationSystem}, a standard truck model also studied in recent work \citep{mcarthur2016truck}. The model under study is a rectangular bluff body with a reference non-dimensional width $W=1$, length $L = 7.647W$, and rounded leading edges with radius $r=0.118W$. The inflow velocity is uniform with a magnitude of $U_\infty$, and the Reynolds number, defined as $\Rey = U_\infty W / \nu$, where $\nu$ represents the fluid kinematic viscosity, is set to $500$. The rectangular computational domain extends from $(-7W, 23W)$ in the streamwise direction ($x$) and $(-7.5W, 7.5W)$ in the transverse direction ($y$), with the front of the bluff body placed at $x = 0$. The simulation domain employs a hybrid mesh, structured near the wall and unstructured elsewhere, encompassing around $168,000$ cells. The time is scaled using the convective time $t_c=W/U_\infty$. The \ac{DNS} is executed in OpenFOAM, using the Gym-preCICE~\citep{gymprecice} wrapper for the coupling library~\citep{preCICEv2} and the OpenFOAM adapter~\citep{OpenFOAMpreCICE} to integrate the simulation with the specified control sequence.
This Reynolds number is chosen as a representative bluff-body wake configuration to develop and validate control-oriented \acp{ROM} before scaling to higher-$\Rey$ turbulent regimes. At $\Rey = 500$ the wake already exhibits the features relevant to this study, namely vortex shedding and sensitivity to base actuation, while the two-dimensional flow assumption remains valid, as verified in advance with a three-dimensional simulation. The compactness--predictability trade-off identified in this work, rooted in the spectral organisation of \ac{POD} versus the broadband character of the \ac{CAE} latent dynamics, is not expected to be specific to this Reynolds number regime; an explicit assessment at higher $\Rey$ and in three-dimensional configurations is left for future work.

\begin{figure}[ht!]
  \centering
    \input{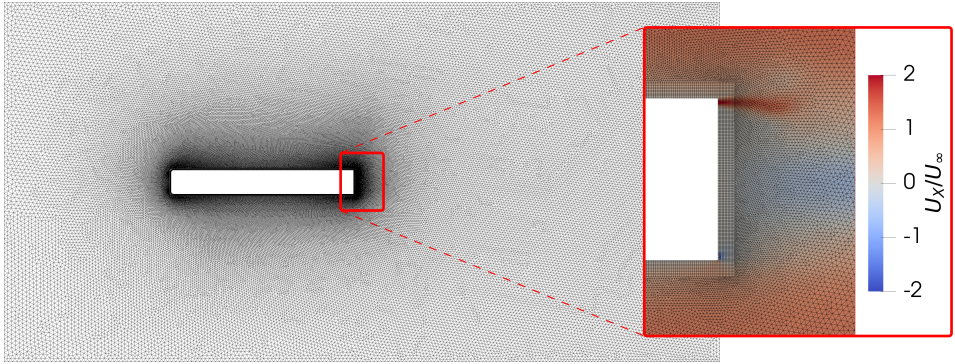}
    \caption{\ac{DNS} mesh and flow details near the base of the truck when jets operate at peak suction/blowing. The symbol $U_x$ in the inset indicates the streamwise velocity component. Figure adapted from~\citep{solera2025framework}.}
    \label{fig:case_jets}
\end{figure}

The control is implemented using two opposing flow jets at the sides of the vehicle base with zero-net mass flow, similar to previous experimental setups~\citep{barros2016coanda, cerutti2020van}. In this case, the overall mass flow rate of the two jets combined is forced to be zero at each instant of the simulation. The jets, each with a width of $w_{jet}=0.05W$, exhibit parabolic velocity profiles with a maximum mean velocity of $1.5U_\infty$. The dataset comprises a time series of velocity fields, with a time increment of $\Delta t = t_c/5 = W/{5U_\infty}$. Figure~\ref{fig:case_jets} illustrates the global mesh and a flow sample around the jet region.

The control input for the simplified truck configuration was generated using a filtered-noise excitation strategy to ensure a rich broadband excitation. First, a discrete white Gaussian noise sequence was generated:
\begin{equation}
  \xi_n \sim \mathcal{N}(0,1)
  \label{eq:noise_gen}
\end{equation}
where $n \in \mathbb{N}_0$ represents the discrete time step index. This index corresponds to the physical time $t_n = n\Delta t$, with a control update interval of $\Delta t = 0.2 t_c$.

To target the relevant flow dynamics, this stochastic sequence was processed through a fourth-order Butterworth bandpass filter, denoted by the operator $\mathcal{B}_4$:
\begin{equation}
  \tilde{\xi}_n = \mathcal{B}_4 \left( \xi_n; f_{\text{low}}, f_{\text{high}} \right)
  \label{eq:noise_filter}
\end{equation}
The filter cut-off frequencies were set to $f_{\text{low}} = 0.05 t_c^{-1}$ and $f_{\text{high}} = 0.5 t_c^{-1}$ to encompass the wake's natural shedding frequency of $f_{sh} \approx 0.2 t_c^{-1}$. The filter was applied in a zero-phase (forward-backward) configuration to prevent distortion.

Finally, the filtered sequence $\tilde{\xi}_n$ was bounded to reflect the physical limits of the synthetic jets and scaled to determine the applied jet flow rate, $b(t_n)$:
\begin{equation}
  b(t_n) = \dot{m}_{\text{max}} \min \left( 1, \max \left( -1, \tilde{\xi}_n \right) \right)
  \label{eq:control_signal}
\end{equation}
where the maximum volumetric flow rate capacity of the actuator is $\dot{m}_{\text{max}} = 0.075 U_{\infty} W$. During the simulation, the environment linearly interpolated the commanded values between the discrete points $t_n$ and $t_{n+1}$ to provide a continuous forcing signal to the \ac{DNS} solver.

Following the simulation, a data preprocessing pipeline was executed to convert the raw OpenFOAM output into a structured dataset. The velocity fields (with $u, v$ being respectively the streamwise and crosswise velocity components) were first interpolated onto a uniform Cartesian grid of $128 \times 256$ points using cubic interpolation. This grid spans the domain from $x/W = -6$ to $x/W = 22$ and $y/W = -7$ to $y/W = 7$. A mask was applied to set to zero the velocity in the grid points inside the truck body. The velocity fields $\boldsymbol{U}(\boldsymbol{x}, t)$ were then normalised by subtracting the temporal mean field, $\boldsymbol{\bar{U}}(\boldsymbol{x})$, and dividing the components $u$ and $v$ by their respective global scalar standard deviations ($\sigma_u$, $\sigma_v$), calculated throughout the entire dataset. In parallel, the actual jet flow rate was extracted from post-processing files and time-synchronised with the velocity snapshots. The final $10\%$ snapshots of the long sequence were used for testing, with the remaining used for training.

\subsubsection{Fluidic pinball}

\begin{figure}[ht!]
\centering
\begin{tikzpicture}[
    scale=1, 
    annot/.style={-{Stealth[length=2mm, width=1.5mm]}},
    every node/.style={font=\sffamily} 
]

\def\cylR{1} 
\pgfmathsetmacro{\yTop}{1.5*\cylR}
\pgfmathsetmacro{\xLeft}{-sqrt(6.75)*\cylR} 

\coordinate (C_L) at (\xLeft, 0);
\coordinate (C_T) at (0, \yTop);
\coordinate (C_B) at (0, -\yTop);

\draw[line width=0.6mm] (C_L) circle (\cylR); 
\draw[line width=0.6mm] (C_T) circle (\cylR); 
\draw[line width=0.6mm] (C_B) circle (\cylR); 

\draw[-{Stealth[length=1.5mm]}] (C_T) ++(-45:\cylR+0.15) arc (-45:135:\cylR+0.15);
\draw[-{Stealth[length=1.5mm]}] (C_B) ++(-45:\cylR+0.15) arc (-45:135:\cylR+0.15);
\draw[-{Stealth[length=1.5mm]}] (C_L) ++(-45:\cylR+0.15) arc (-45:135:\cylR+0.15);

\draw ($(C_T) + (150: \cylR)$) -- node[below=1pt] {} (C_T);
\draw[annot] ($(C_T) + (150: \cylR+1)$) -- node[above, sloped] {$R$} ($(C_T) + (150: \cylR)$);

\draw [opacity=0.7] (C_T) -- ++(1.5, 0); 
\draw [opacity=0.7] (C_B) -- ++(1.5, 0); 
\draw[{Latex[length=1.5mm]}-{Latex[length=1.5mm]}] 
    (1.4, -\yTop) -- node[right, xshift=2mm] {$3R$} (1.4, \yTop);

\draw [opacity=0.7] (C_L) -- ++(-0.75, -1.5*0.866);
\draw [opacity=0.7] (C_B) -- ++(-0.75, -1.5*0.866);
\draw[{Latex[length=1.5mm]}-{Latex[length=1.5mm]}] 
    ($(C_L)+(-0.7, -1.4*0.866)$) -- node[below, sloped, yshift=-1mm] {$3R$} ($(C_B)+(-0.7, -1.4*0.866)$);

\draw[dash dot, opacity=0.7] (\xLeft - 1.25*\cylR, 0) -- (\cylR, 0); 
\draw[dashed, opacity=0.7] (0, \yTop + \cylR*1.25) -- (0, -\yTop - \cylR*1.25);

\draw[-{Stealth[length=4mm, width=3mm]}, line width=1.25pt] 
    (\xLeft - 3.5*\cylR, 0) -- (\xLeft - 2.5*\cylR, 0);
\node[above=2pt] at (\xLeft - 3*\cylR, 0) {$U_{\infty}$};

\end{tikzpicture}
\caption{Schematic of the fluidic pinball flow configuration.}
\label{fig:PinballSchematic}
\end{figure}
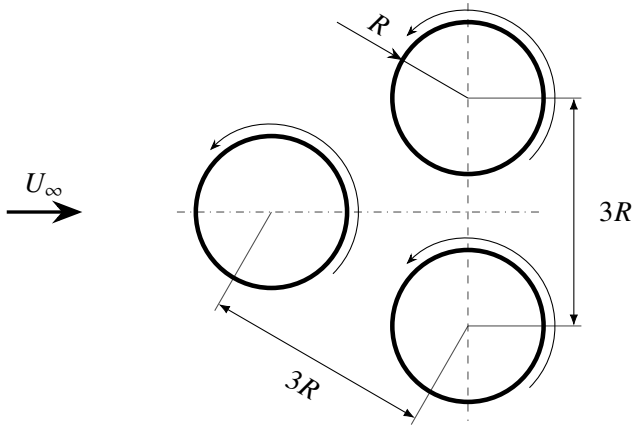

The second test case is the Fluidic pinball, a canonical 2D setup often used for complex multi-input multi-output (MIMO) control problems~\citep{deng2020}, schematised in Figure~\ref{fig:PinballSchematic}. The geometry consists of three cylinders of equal diameter $D=2R$, with their centres located at the vertices of an equilateral triangle. The centre-to-centre side length of the triangle is $3R$. The origin of the reference system is located in the midpoint between the two centres of the downstream cylinders, with the $x$ axis pointing in the direction and the sense of the free-stream velocity $U_\infty$ and the $y$ axis perpendicular to it in the transverse direction. The rectangular computational domain extends from $(-5D, 15D)$ in the streamwise ($x$) direction and $(-5D, 5D)$ in the transverse ($y$) direction.

The dataset for this configuration was generated with a solver based on an implicit time integration finite element method. The Reynolds number is set to $\Rey=150$, higher than the transition point $\Rey \simeq 115$ from a quasi-periodic asymmetric regime to chaotic flow~\citep{deng2020, deng2025pinball}. This regime exhibits a dominant vortex shedding with random upward and downward switches in the centre jet, remaining as a challenging dynamical system. The time is scaled using the convective time $t_{c}=D/U_\infty$. The resulting dataset contains the 2D velocity fields ($u, v$) for each time step, normalised with the free-stream velocity $U_\infty$. Control is applied by changing the tangential speed of the three cylinder surfaces, providing three independent control parameters. Notice that the input speeds are normalised with $U_\infty$ and defined positive when the rotation is set to be counter-clockwise.

The control setup for the fluidic pinball configuration was designed as a multi-input strategy using quasi-stationary step ramps to excite the dynamics of the system. The control inputs are defined by the normalised tangential speeds of the three cylinders, $\mathbf{b}(t) = [b_1(t), b_2(t), b_3(t)]^T$, corresponding to the top, bottom, and forward cylinders, respectively.

To cover the control spectrum, target speed combinations, denoted as $\mathbf{v}_k$, were generated from the $125$ possible permutations of discrete rotational speeds:
\begin{equation}
  \mathbf{v}_k \in \mathcal{V}^3, \quad \text{where} \quad \mathcal{V} = \{-1, -0.5, 0, 0.5, 1\}
  \label{eq:pinball_combinations}
\end{equation}

For a given target step $k$, each cylinder $i \in \{1, 2, 3\}$ maintains its target speed $v_{k,i}$ for a constant hold duration of $T_{\text{hold},i}$. Following the hold phase, the speed transitions linearly to the subsequent target value $v_{k+1,i}$ over a cylinder-specific ramp duration $T_{\text{ramp},i}$.

Mathematically, the control signal for cylinder $i$ during the transition to the $(k+1)$-th step and its subsequent hold phase can be described by the piecewise function:
\begin{equation}
  b_i(t) =
  \begin{cases} 
  v_{k,i} + \frac{v_{k+1,i} - v_{k,i}}{T_{\text{ramp}, i}} (t - t_{k,i}), & \text{if } t_{k,i} \le t < t_{k,i} + T_{\text{ramp}, i} \\ 
  v_{k+1,i}, & \text{if } t_{k,i} + T_{\text{ramp}, i} \le t \le t_{k+1,i} 
  \end{cases}
  \label{eq:pinball_control}
\end{equation}

where $t_{k,i}$ marks the beginning of the $k$-th ramp for cylinder $i$, and $t_{k+1,i} = t_{k,i} + T_{\text{ramp}, i} + T_{\text{hold}, i}$ is the beginning of the next cycle. The transition durations vary for each actuator, defined as $T_{\text{ramp}, 1} = 25\,t_c$ for the top cylinder, $T_{\text{ramp}, 2} = 40\,t_c$ for the bottom cylinder, and $T_{\text{ramp}, 3} = 120\,t_c$ for the forward cylinder. Similarly, the hold durations are also specific to each cylinder, i.e. $T_{\text{hold}, 1} = 30\,t_c$, $T_{\text{hold}, 2} = 235\,t_c$ and $T_{\text{hold}, 3} = 1250\,t_c$.

This dataset contains a total of 70,000 snapshots, captured with a time spacing of 0.1 convective times between each snapshot. The final 20,000 snapshots are reserved for testing, with the first 50,000 used for training. Following the simulation, each snapshot of the velocity field was preprocessed and interpolated onto a regular grid, resulting in a final resolution of $96 \times 192$ pixels. As with the previous flow configuration, a zero velocity mask is applied to the inner body regions. Both \ac{POD} and \ac{CAE} operate on the fluctuating velocity components $\boldsymbol{u}(\boldsymbol{x}, t) = \boldsymbol{U}(\boldsymbol{x}, t) - \bar{\boldsymbol{U}}(\boldsymbol{x})$, normalised by their respective global standard deviations.

\begin{table}
\centering
\caption{Summary of the datasets for the two flow configurations.}
\label{tab:dataset_summary}
\begin{tabular}{lcc}
\toprule
Parameter & Simplified truck & Fluidic pinball \\
\midrule
Reynolds number & 500 & 150 \\
Control type & Single-input (synthetic jets) & Multi-input (rotating cylinders) \\
Control signal & Filtered random & Quasi-stationary step ramps \\
Control dimension & 1 & 3 \\
Total snapshots & 50,000 & 70,000 \\
Test snapshots & 5,000 & 20,000 \\
Time step ($\Delta t$) & $0.2 t_c$ & $0.1 t_{c}$ \\
Spatial resolution ($N_y \times N_x$) & $128 \times 256$ & $96 \times 192$ \\
\bottomrule
\end{tabular}
\end{table}

\subsection{Dimensionality reduction methods}

The datasets consist of long time-series of full-field velocity field snapshots under different control actions. Attempting to directly predict the temporal evolution of the entire high-dimensional spatial field, $\boldsymbol{U}(\boldsymbol{x}, t)$, is computationally intractable for real-time control applications. The core strategy is therefore to first reduce the high spatial dimensionality of each snapshot into a low-dimensional latent-space representation, $\boldsymbol{z}(t) \in \mathbb{R}^d$, where $d \ll N_p$ is the latent dimension, and $N_p$ the number of grid points. The subsequent task, detailed in \S~\ref{subsec:predictors}, is to build a predictive model that can forecast the temporal evolution of this much smaller state vector. In this work, we implement and systematically compare two canonical families of methods for the spatial dimensionality reduction task: the classical, energy-optimal linear \ac{POD}, and nonlinear deep \acp{CAE} following previous work such as ~\cite{fukagata2025compressing} and \cite{ solera2024bvae}. Within the nonlinear family, we additionally consider two regularised variational variants, a \ac{bVAE}  and a \ac{dklVAE}, to assess whether latent-space regularisation modifies the resulting latent dynamics. The following sections detail the implementation of each model.

To compare the performance of \ac{CAE} and \ac{POD} methods, we define the energy percentage $E$ metric, following ~\cite{solera2024bvae}, \cite{eivazi2022vae}, a metric of the energy captured by the low-order reconstruction as:
\begin{equation}
    E = \left( 1 - \left\langle
\frac{\sum_{j=1}^{N_p}\sum_{i=1}^{N_c}\left(u_{i,j}-\tilde{u}_{i,j}\right)^2}
         {\sum_{j=1}^{N_p}\sum_{i=1}^{N_c} u_{i,j}^2}
    \right\rangle \right) \times 100\,\%,
    \label{ec:E}
\end{equation}

\noindent where $\langle \cdot \rangle$ indicates ensemble averaging in time, $N_p$ is the number of grid points, $N_c$ the number of velocity components, $u_{i,j}$ denotes the $i$-th component of the reference fluctuating velocity at grid point $j$ and $\tilde{u}_{i,j}$ its low-order reconstruction. The same metric is evaluated at two stages of the framework: on the decoded snapshot to assess the compression quality of \ac{POD} or \ac{CAE}, and on the decoded prediction at a horizon $\tau$ (in convective time $t_c$ units) to assess the end-to-end prediction accuracy. When the time average $\langle \cdot \rangle$ is omitted, the same error ratio evaluated for a single snapshot is referred to as the \textit{instantaneous} accuracy; the pooled distributions reported in \S~\ref{sec:results} are built from these instantaneous values.

\subsubsection{Proper orthogonal decomposition}

\ac{POD} is employed as a linear baseline to compare against the nonlinear \ac{CAE}. \ac{POD} provides an optimal linear basis to represent a dataset in terms of the $L_2$ norm, which can be interpreted in this case as turbulent kinetic energy. We first decompose the velocity field $\boldsymbol{U}(\boldsymbol{x},t)$ (which contains both $u$ and $v$ components) as:

\begin{equation}
\boldsymbol{U}(\boldsymbol{x},t) = \boldsymbol{\bar{U}}(\boldsymbol{x}) + \boldsymbol{u}(\boldsymbol{x},t),
\label{Eq:VelDec}
\end{equation}

\noindent where $\boldsymbol{\bar{U}}(\boldsymbol{x})$ is the velocity field averaged over time and $\boldsymbol{u}(\boldsymbol{x},t)$ is the fluctuating component. The fluctuating velocity field can be approximated as a linear combination of orthonormal spatial basis functions $\boldsymbol{\phi}_i(\boldsymbol{x})$, the \ac{POD} modes:

\begin{equation}
\boldsymbol{u}(\boldsymbol{x},t) \approx \sum_{i=1}^{d} a_i(t) \boldsymbol{\phi}_i(\boldsymbol{x}),
\end{equation}

\noindent where $a_i(t)$ are the time-dependent modal coefficients  and $d$ is the number of modes retained for reconstruction.

To find the modes, the snapshot matrix $\boldsymbol{S}$ is assembled from the training data, where each row is a flattened velocity snapshot $\boldsymbol{u}(\boldsymbol{x}, t_j)$. Let $N_t$ be the number of training snapshots and $N_p$ be the number of grid points. The matrix $\boldsymbol{S}$ thus has dimensions $N_t \times (N_pN_c)$, with $N_c$ being the number of velocity components. Following the method of snapshots~\citep{sirovich1987turbulence}, the spatial modes $\boldsymbol{\phi}_i$ are found by solving the eigenvalue problem for the spatial covariance matrix $\boldsymbol{C}$:

\begin{equation}
\boldsymbol{C} = \frac{1}{N_t - 1} \boldsymbol{S}^T \boldsymbol{S},
\end{equation}

\noindent where $\boldsymbol{C}$ is an $N_pN_c \times N_pN_c$ matrix. The eigenvectors of $\boldsymbol{C}$, sorted by their corresponding eigenvalues $\lambda_i$, are the \ac{POD} modes $\boldsymbol{\phi}_i$. The eigenvalues represent the kinetic energy content of each mode. Given the high dimensionality of $\boldsymbol{C}$, for computational efficiency we compute only the leading $d$ eigenvectors using a randomised SVD algorithm.

Once the spatial modes $\boldsymbol{\Phi}$ (a matrix of size $N_p N_c \times d$) are known, the temporal modes $\boldsymbol{A}$ (a matrix of size $N_t \times d$) are obtained by projecting the snapshot matrix onto the modes:

\begin{equation}
\boldsymbol{A} = \boldsymbol{S} \boldsymbol{\Phi}.
\end{equation}

\noindent This matrix $\boldsymbol{A}$ of temporal coefficients forms the latent-space representation for \ac{POD}-based models, analogous to the latent vector $\boldsymbol{z}$ of the \ac{CAE}.

\subsubsection{Convolutional autoencoder}

An \ac{AE} is a neural network architecture designed for dimensionality reduction~\citep{hinton2006ae}, commonly used for feature learning and data compression. It consists of two main components: an encoder ($\mathcal{E}$) and a decoder ($\mathcal{D}$). The encoder maps the high-dimensional input $\boldsymbol{u}$ to a low-dimensional latent-space vector $\boldsymbol{z} = \mathcal{E}(\boldsymbol{u})$. The decoder then reconstructs the input from this latent vector, $\tilde{\boldsymbol{u}}=\mathcal{D}(\boldsymbol{z})$, aiming to obtain a reconstruction $\tilde{\boldsymbol{u}}$ as close to the original  $\boldsymbol{u}$ as possible.

The encoder and decoder networks are trained simultaneously by gradient descent and backpropagation using the Adam algorithm~\citep{adam2014method}, with the hyperparameters detailed in table~\ref{tab:training_params}. The training process minimises a loss function that measures the discrepancy between the input $\boldsymbol{u}$ and the reconstruction $\tilde{\boldsymbol{u}}$, defined as the squared $L_2$ norm of the reconstruction error:
$\mathcal{L}_{rec} = \left\| \boldsymbol{u} - \tilde{\boldsymbol{u}} \right\|^2_2$.
Once trained, the encoder is used to generate the temporal evolution of the latent space by applying it to each time step: $\boldsymbol{z}_t = \mathcal{E}(\boldsymbol{u}_t)$.

The spatial nature of flow field data guides the choice of a \ac{CNN} to build the encoder and decoder networks, since spatial patterns are encoded more naturally by convolutional layers~\citep{lecun1998cnn}. In the encoder, we use five or six convolutional layers (for the pinball and truck case, respectively) with a stride of two so that each layer halves the spatial dimension. This spatial reduction allows the subsequent layers to capture information at larger scales in the input flow data. As the spatial dimensions decrease, the count of filters in each layer increases to maintain information of the flow. After the last convolutional layer, the spatial information is flattened, and a fully connected layer is added to combine the information. Finally, a single linear layer with $d$ units outputs the latent vector $\boldsymbol{z}$.

The decoder model is configured as a nearly symmetric network relative to the encoder. The latent vector $\boldsymbol{z}$ is input into a fully-connected layer, with its output reshaped to match the last convolutional layer of the encoder. Subsequently, five to six transposed convolution layers are applied to incrementally expand the spatial dimension while reducing the number of filters. The last transposed convolution layer, using two filters, generates the output channels for the $u$ and $v$ velocity components. The activation function is the exponential linear unit~\citep{clevert2015elu} for all layers except the last, where the activation is linear.

\begin{table}
\centering
\caption{ \ac{CAE}, \ac{bVAE} and \ac{dklVAE} training hyperparameters for each case. $\beta$ applies to the \ac{bVAE}, and the loss weights $\lambda_{\text{MI}}$, $\lambda_{\text{TC}}$, $\lambda_{\text{Dim}}$ to the \ac{dklVAE}.}
\label{tab:training_params}
\begin{tabular}{lcc}
\toprule
Hyperparameter & Simplified truck & Fluidic pinball \\
\midrule
Latent Dimension ($d$) & 8 & 10\\
$\beta$ (\ac{bVAE}) & 0.0025 & 0.005\\
$\lambda_{\text{MI}}$ (\ac{dklVAE}) & $1.6 \times 10^{-4}$ & $1 \times 10^{-4}$\\
$\lambda_{\text{TC}}$ (\ac{dklVAE}) & $5 \times 10^{-4}$ & $5 \times 10^{-3}$\\
$\lambda_{\text{Dim}}$ (\ac{dklVAE}) & $1.6 \times 10^{-4}$ & $1 \times 10^{-4}$\\
Batch Size & 1024 & 256 \\
Learning Rate & $1 \times 10^{-3}$ & $1 \times 10^{-3}$ \\
Epochs & 200 & 300 \\
Test snapshots & 5,000 & 20,000 \\
\bottomrule
\end{tabular}
\end{table}

\subsubsection{$\beta$-Variational autoencoder}

In addition to the deterministic \ac{CAE}, we also include a \ac{bVAE}~\citep{solera2024bvae} as a regularised nonlinear baseline. The \ac{bVAE} shares the same convolutional encoder-decoder backbone as the \ac{CAE} described above, but uses a probabilistic latent representation~\citep{bVAE_higgins2017betavae}. The encoder outputs the parameters of a Gaussian distribution in latent space, $\boldsymbol{\mu}(\boldsymbol{u})$ and $\boldsymbol{\sigma}(\boldsymbol{u})$, from which the latent vector is sampled via the reparameterisation trick, $\boldsymbol{z} = \boldsymbol{\mu} + \boldsymbol{\sigma} \odot \boldsymbol{\epsilon}$, with $\boldsymbol{\epsilon} \sim \mathcal{N}(\boldsymbol{0}, \boldsymbol{I})$. The decoder reconstructs $\tilde{\boldsymbol{u}}$ from $\boldsymbol{z}$. This probabilistic structure allows us to assess whether constraining the geometry of the nonlinear latent manifold can mitigate the loss of long-horizon predictability observed for the \ac{CAE}.

The training loss combines the reconstruction error with the Kullback-Leibler divergence $D_{\text{KL}}$, which measures the discrepancy between the encoded latent distribution and a unit Gaussian prior:

\begin{equation}
\mathcal{L}_{\beta\text{-VAE}} = \| \boldsymbol{u} - \tilde{\boldsymbol{u}}  \|^2_2 + \beta \, D_{\text{KL}} , 
\end{equation}

\noindent where $\beta$ controls the strength of the regularisation. Larger values of $\beta$ promote more disentangled and statistically independent latent coordinates at the cost of reconstruction fidelity~\citep{solera2024bvae}. The parameter $\beta$ is tuned to increase in statistical independence relative to the deterministic \ac{CAE} without significantly increasing reconstruction error. The \ac{bVAE} is trained with the same latent dimension and remaining hyperparameters as the \ac{CAE}, reported in Table~\ref{tab:training_params}.

\subsubsection{Decomposed-KL variational autoencoder}

A limitation of the \ac{bVAE} is that the single coefficient $\beta$ weights the whole Kullback--Leibler term, which couples several effects: promoting independence between latent coordinates also forces each marginal towards the prior, and an excessive penalty can reduce the information capacity of the latent space. To assess whether a more selective regularisation modifies the latent dynamics, we consider a \ac{dklVAE}, in which the divergence is split into three independently weighted contributions following the decomposition of~\cite{chen2018isolating}, recently applied to manifold learning of fluid flows by~\cite{wang2026information}:

\begin{equation}
\mathcal{L}_{\text{DKL-VAE}} = \| \boldsymbol{u} - \tilde{\boldsymbol{u}} \|^2_2
+ \lambda_{\text{MI}}\, \mathcal{I}
+ \lambda_{\text{TC}}\, \mathcal{T}
+ \lambda_{\text{Dim}}\, \mathcal{K},
\end{equation}

\noindent where $\mathcal{I}$ is the index-code mutual information between the snapshot index and the latent code, $\mathcal{T}$ is the total correlation that measures the statistical dependence among latent coordinates, and $\mathcal{K}$ is the dimension-wise divergence between each marginal latent distribution and the prior. These three terms sum to the standard Kullback--Leibler divergence, so the \ac{bVAE} is recovered when $\lambda_{\text{MI}}=\lambda_{\text{TC}}=\lambda_{\text{Dim}}=\beta$. Increasing $\lambda_{\text{TC}}$ alone encourages independent latent coordinates without tightening the matching of each coordinate to the prior, thereby decoupling disentanglement from prior matching. The aggregated posterior required by these terms is estimated through minibatch stratified sampling~\citep{chen2018isolating}. We evaluate the \ac{dklVAE} on the simplified truck and fluidic pinball cases as an alternative regularisation strategy, using $\lambda_{\text{MI}}=\lambda_{\text{Dim}}=1.6\times10^{-4}$ and $\lambda_{\text{TC}}=5\times10^{-4}$ for the truck, and $\lambda_{\text{MI}}=\lambda_{\text{Dim}}=1\times10^{-4}$ and $\lambda_{\text{TC}}=5\times10^{-3}$ for the pinball (Table~\ref{tab:training_params}), keeping the same convolutional backbone, latent dimension and remaining hyperparameters as the \ac{CAE} and \ac{bVAE}.

\subsection{\label{subsec:predictors}Time-series prediction models}

Once high-dimensional spatial data $\boldsymbol{U}(\boldsymbol{x}, t)$ is compressed into a low-dimensional latent-space time-series $\boldsymbol{z}(t)$, the next task is to build a predictive model that can predict the evolution of this latent state. The goal is to learn a function $f$ that predicts a future sequence of $H$ latent states (the \textit{horizon}), given a sequence of $L$ past latent states and their corresponding control inputs (the \textit{lookback}). Critically, the control action $\boldsymbol{b}_t$ applied at time $t$ influences the state at $t+1$. Therefore, a causal model must predict the discrete state sequence $[\hat{\boldsymbol{z}}_{t+1}, \dots, \hat{\boldsymbol{z}}_{t+H}]$ using the control sequence $[\boldsymbol{b}_{t-L}, \dots, \boldsymbol{b}_t, \dots \boldsymbol{b}_{t+H-1}]$:

\begin{equation}
    [\hat{\boldsymbol{z}}_{t+1}, \dots, \hat{\boldsymbol{z}}_{t+H}] = 
    f([\boldsymbol{z}_{t-L+1}, \dots, \boldsymbol{z}_t],
      \boldsymbol{b}_{t-L}, \dots, 
      [\boldsymbol{b}_{t}, \dots, \boldsymbol{b}_{t+H-1}]),
\end{equation}
\noindent where $\hat{\boldsymbol{z}}$ denotes the predicted state.

From a closure-modelling standpoint, projection-based \acp{ROM} are not in general closed~\citep{Noack2003CylinderWake}. An in-principle infinite Mori--Zwanzig memory kernel encodes the action of unresolved scales on the resolved coordinates~\citep{Menier2023CDROM,Gupta2025MZKoopman}. We adopt as a working hypothesis a finite-memory approximation in which this kernel is truncated to a window of length $L$ and represented by the learned predictor, which absorbs both the resolved-mode contribution and a deterministic, data-driven surrogate of the orthogonal-dynamics term. Under this hypothesis, latent representations that better separate resolved and unresolved scales are expected to render the truncated closure a closer approximation; the empirical consequences are revisited in \S\ref{subsec:dynamics_analysis}.

We restrict the temporal modelling stage to \ac{LSTM}-based predictors~\citep{Hochreiter1997lstm} for two reasons. First, \acp{LSTM} are a well-established baseline for nonlinear time-series modelling in fluids and \ac{ROM} settings \citep{solera2025framework}, and they provide sufficient expressive power to test our main question, namely how the choice of spatial encoder affects predictability in the latent space. Second, \acp{LSTM} offer a favourable accuracy-complexity balance: they are comparatively lightweight, stable to train on moderate datasets, and amenable to real-time deployment on embedded hardware. By keeping the predictor family fixed, we ensure a controlled comparison in which differences in forecasting performance can be attributed primarily to the latent representations produced by \ac{POD} versus \acp{CAE}.

\begin{figure}[!ht]
    \centering
    \includegraphics[width=0.9\linewidth]{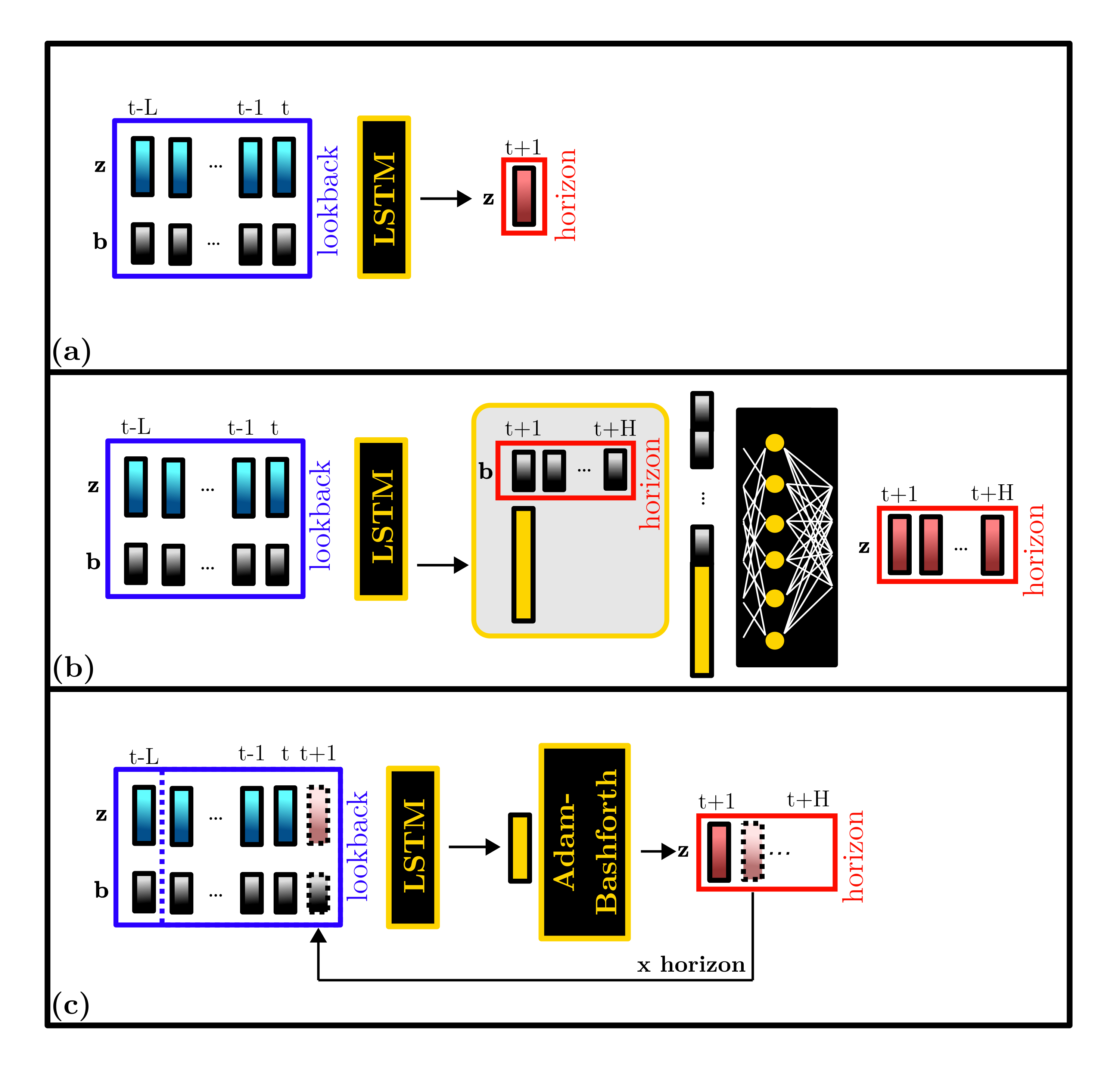}
    \caption{Schematic of the three \ac{LSTM}-based prediction architectures tested. For all architectures, the latent space (in blue) and control space (in gray) are fed as input throughout the entire lookback period up to the current time step $t$. (Top) Single-step model , consisting of an \ac{LSTM} layer that predicts the immediate next time step of the latent space $\boldsymbol{z}_{t+1}$ (in red). (Middle) Direct sequence model. The latter predicts the entire horizon sequence of the latent space at once by combining a latent context vector (yellow bar) as the output of the \ac{LSTM} layer with the horizon sequence of the control space through an \ac{MLP} layer. (Bottom) Derivative-based model. The \ac{LSTM} predicts the time derivative of the latent state $\dot{\boldsymbol{z}}_t$. An Adams-Bashforth method is used to propagate the solution to $\boldsymbol{z}_{t+1}$ with derivative information. The process is repeated up to time instant $t+H$ by feeding the predictions into the advanced lookback sequence.}
    \label{fig:4_LSTMs_scheme}
\end{figure}

The input to the \ac{LSTM} at each step in the lookback sequence is the concatenation of the latent state and the control signal at that time. In this study, we implement and compare three distinct predictive \ac{LSTM} architectures, illustrated in Figure~\ref{fig:4_LSTMs_scheme}.

\begin{itemize}
    \item \textbf{Single-step:} This is a standard autoregressive model, corresponding to the top schematic in Figure~\ref{fig:4_LSTMs_scheme}. The \ac{LSTM} takes the lookback sequence (up to $\boldsymbol{z}_t$ and $\boldsymbol{b}_t$) and predicts only the single next latent state, $\hat{\boldsymbol{z}}_{t+1}$. To generate long-term predictions, this output is combined with the next control input $\boldsymbol{b}_{t+1}$ and fed back into the model to predict $\hat{\boldsymbol{z}}_{t+2}$. This process is repeated recursively, thus potentially suffering from error propagation.

    \item \textbf{Direct sequence:} This model, shown in the middle of Figure~\ref{fig:4_LSTMs_scheme}, is a sequence-to-sequence architecture. It takes the lookback sequence as input to the \ac{LSTM} core. The final hidden state of the \ac{LSTM} (which encodes information up to time $t$) is then concatenated with the future control sequence that will modify the future states, $[\boldsymbol{b}_t, \dots, \boldsymbol{b}_{t+H-1}]$, and passes through a \ac{MLP}. This network directly predicts the entire future sequence of latent states $[\hat{\boldsymbol{z}}_{t+1}, \dots, \hat{\boldsymbol{z}}_{t+H}]$ in a single forward pass, avoiding the recursive error accumulation of the single-step model. While this requirement for the complete future control sequence makes it unsuitable for standard reactive controllers, it is the ideal structure for \ac{MPC}, where this exact future control sequence is the variable being optimised.

    \item \textbf{Derivative-based:} This hybrid architecture, shown at the bottom of Figure~\ref{fig:4_LSTMs_scheme}, embeds a numerical time integrator within the prediction loop as described in \cite{yang2025adameul}. Instead of directly predicting the future state $\hat{\boldsymbol{z}}_{t+1}$, the \ac{LSTM} is trained to predict the time derivative of the latent state $\dot{\boldsymbol{z}}$, based on the inputs at time $t$, $\dot{\boldsymbol{z}}_t=g(\boldsymbol{z}_t,\boldsymbol{b}_t)$. The future state is then obtained using an explicit time-stepping scheme. The model uses the second-order Adams-Bashforth method:
    \begin{equation}
        \hat{\boldsymbol{z}}_{t+1} = \hat{\boldsymbol{z}}_t + \Delta t \left( \frac{3}{2}\dot{\boldsymbol{z}}_t - \frac{1}{2}\dot{\boldsymbol{z}}_{t-1} \right),
    \end{equation}
    \noindent where the \ac{LSTM} provides the derivative $\dot{\boldsymbol{z}}_t$ at each step. $\dot{\boldsymbol{z}}_{t-1}$ is initialised by estimating the initial derivative using a backward finite difference of the last two latent states in the lookback window. This imposes a physical structure on the model, which can improve stability for long-horizon predictions.
\end{itemize}

All three architectures are trained by minimising a loss function, $\mathcal{L}_{\text{pred}}$, based on the $L_2$ norm between the predicted latent-state sequence and the ground truth sequence over the prediction horizon $H$:
\begin{equation}
\mathcal{L}_{\text{pred}} = \frac{1}{H} \sum_{i=1}^{H} \left\| \hat{\boldsymbol{z}}_{t+i} - \boldsymbol{z}_{t+i} \right\|^2_2.
\end{equation}

\noindent While \texttt{Single-step} and \texttt{Direct sequence} models are trained using this standard $L_2$ loss, the \texttt{Derivative-based} architecture employs a specific loss, also introduced in~\cite{yang2025adameul}, which weights only the $L_2$ norm of the first and last steps in the horizon. This was found to improve the long-term stability of the derivative-based integration. A summary of the training parameters is shown in table~\ref{tab:lstm_params_truck}. We note that the hyperparameters have been tuned via a manual hyperparameter search on the test set metrics, specifically for each predictor architecture, in order to compare each architecture in its optimal configuration. 

\begin{table}
\centering
\caption{\ac{LSTM} training hyperparameters.}
\label{tab:lstm_params_truck}
\begin{tabular}{lcccccccc}
\toprule
Model & $d$ & $L$ & $H$ & \ac{LSTM} lay. & \ac{LSTM} dim. & \ac{MLP} dim. & Dropout \\
\midrule
\multicolumn{8}{c}{\textbf{Simplified 2D truck case}} \\
\midrule
\multicolumn{8}{c}{\ac{POD}-based models} \\
\texttt{Single-step} & 8 & 16 & 1 & 2 & 32 & 32 & 0.0\\
\texttt{Direct sequence} & 8 & 16 & 8 & 2 & 64 & 64 & 0.0\\
\texttt{Derivative-based} & 8 & 4 & 8 & 2 & 64 & 64 & 0.3 \\
\midrule
\multicolumn{8}{c}{\ac{AE}-based models} \\
\texttt{Single-step} & 8 & 8 & 1 & 2 & 128 & 128 & 0.0\\
\texttt{Direct sequence} & 8 & 8 & 2 & 2 & 128 & 128 & 0.0\\
\texttt{Derivative-based} & 8 & 8 & 2 & 2 & 128 & 128 & 0.0 \\
\midrule
\multicolumn{8}{c}{\textbf{Fluidic pinball case}} \\
\midrule
\multicolumn{8}{c}{\ac{POD}-based models} \\
\texttt{Single-step} & 22 & 64 & 1 & 1 & 16 & 16 & 0.3 \\
\texttt{Direct sequence} & 22 & 32 & 16 & 2 & 128 & 256 & 0.0 \\
\texttt{Derivative-based} & 22 & 8 & 16 & 2 & 64 & 64 & 0.3 \\
\midrule
\multicolumn{8}{c}{\ac{AE}-based models} \\
\texttt{Single-step} & 10 & 64 & 1 & 1 & 16 & 16 & 0.3 \\
\texttt{Direct sequence} & 10 & 32 & 16 & 2 & 128 & 256 & 0.0 \\
\texttt{Derivative-based} & 10 & 8 & 16 & 2 & 64 & 64 & 0.3 \\
\bottomrule
\end{tabular}
\end{table}

\section{\label{sec:results}Results}

This section presents a systematic comparison of the performance of the eight main model combinations: \ac{POD} and \ac{CAE} encoding, each paired with the three \ac{LSTM} predictors, plus the \ac{bVAE} and \ac{dklVAE} encodings, each paired with the best-performing predictor. We first evaluate the baseline compression efficiency of encoders, followed by an analysis of the long-term predictive accuracy, and conclude with an investigation into the latent-space dynamics to explain the observed performance.

\subsection{\texorpdfstring{Compression methods: \ac{POD}, \ac{CAE}, \ac{bVAE} and \ac{dklVAE}}{Compression methods: POD, CAE, beta-VAE and dklVAE}}

\begin{figure}[ht!]
    \centering
    \includegraphics[width=\linewidth]{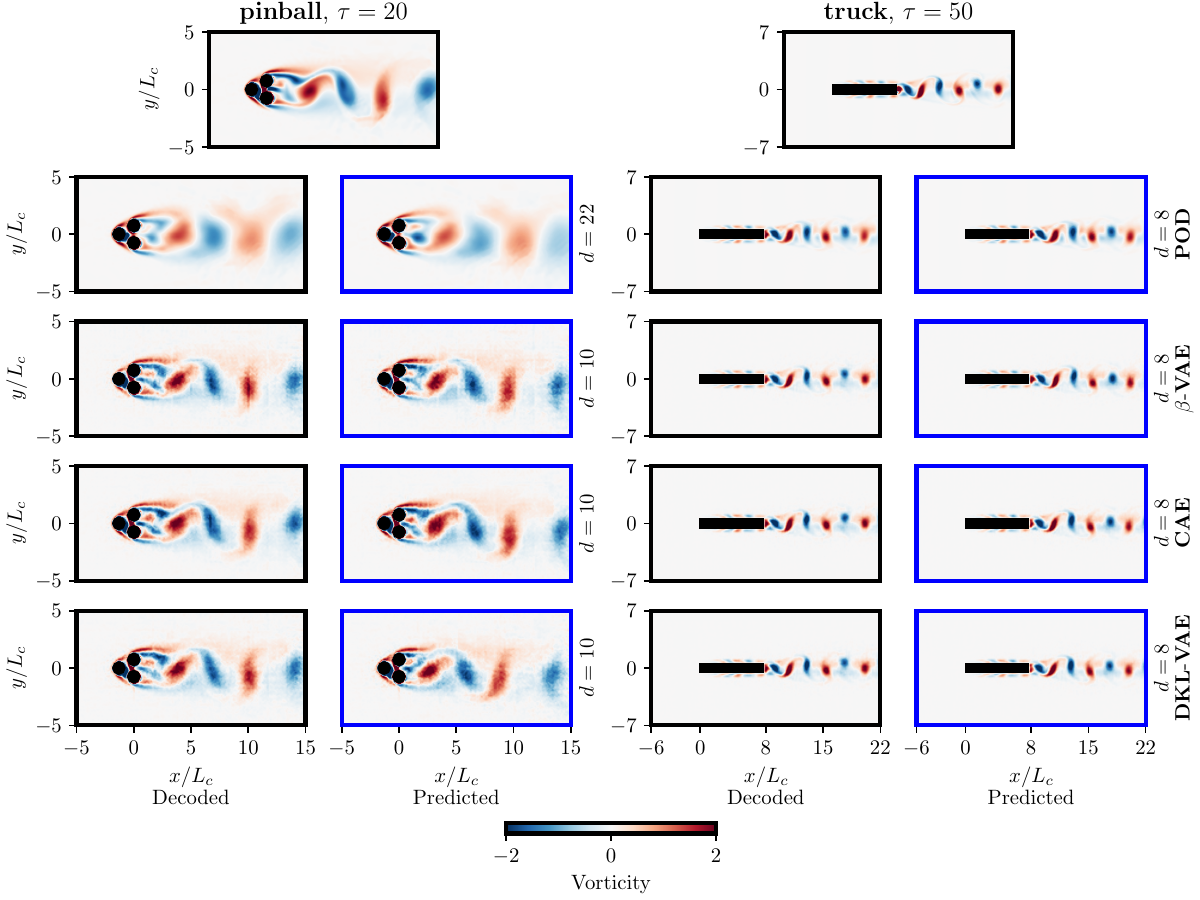}
    \caption{Qualitative comparison of flow field reconstruction and long-term prediction for the fluidic pinball (left, at $\tau = 20$) and the simplified truck (right, at $\tau = 50$). The ``True'' row shows the reference nondimensional vorticity; all rows for a given case are extracted from the same test snapshot. The ``Decoded'' column shows the instantaneous reconstruction from the latent space for the \ac{POD}, \ac{bVAE}, \ac{CAE} and \ac{dklVAE} models, without prediction. The ``Predicted'' column shows the final predicted field decoded from the best \ac{LSTM} models' latent-space predictions after the specified time horizon.}
    \label{fig:fields}
\end{figure}

We first establish the reconstruction accuracy of the four dimensionality reduction methods, independent of the temporal prediction. Figure~\ref{fig:fields} provides a qualitative comparison of the reconstructed velocity fields from \ac{POD}, \ac{CAE}, \ac{bVAE} and \ac{dklVAE} models against the ground truth ``True''. The ``Decoded'' columns show the instantaneous reconstruction of a test snapshot from its latent representation.

In all cases, the latent dimension selection criterion is such that the minimum number of coordinates leads to at least $95\%$ of the original variance of the data for the truck and $90\%$ for the fluidic pinball configuration. These thresholds refer to the mean reconstruction accuracy $E$ averaged over the entire test set (or the cumulative eigenvalue ratio for \ac{POD}). For the simplified truck case (right column of Figure~\ref{fig:fields}), both models use a latent dimension of $d=8$. A more significant difference is seen in the fluidic pinball case (left column). Here, the \ac{CAE}, \ac{bVAE} and \ac{dklVAE}
 are able to encode, across the entire test set, $90\%$ of the variance with a latent space of dimension 10, compared to \ac{POD} which needs 22.

The nonlinear \acp{AE} are more efficient and powerful compressors, capable of capturing the features of the flow under control actions in a more compact latent representation than the linear \ac{POD}, as observed in previous works~\citep{Murata_Fukami_Fukagata_2020, Fukami2020HCAE, eivazi2022vae, zhu2024cnn}. Notably, the \ac{AE} reconstructions in Figure~\ref{fig:fields} retain less diffuse and higher-wavenumber features than \ac{POD}, while \ac{bVAE} prediction exhibits a visible phase shift in the wake, discussed in \S~\ref{subsec:dynamics_analysis}. As shown next, preserving these details can increase the complexity of the latent dynamics and reduce long-horizon predictability.

\subsection{Prediction accuracy}

\begin{figure}[ht!]
    \centering
    \includegraphics[width=\linewidth]{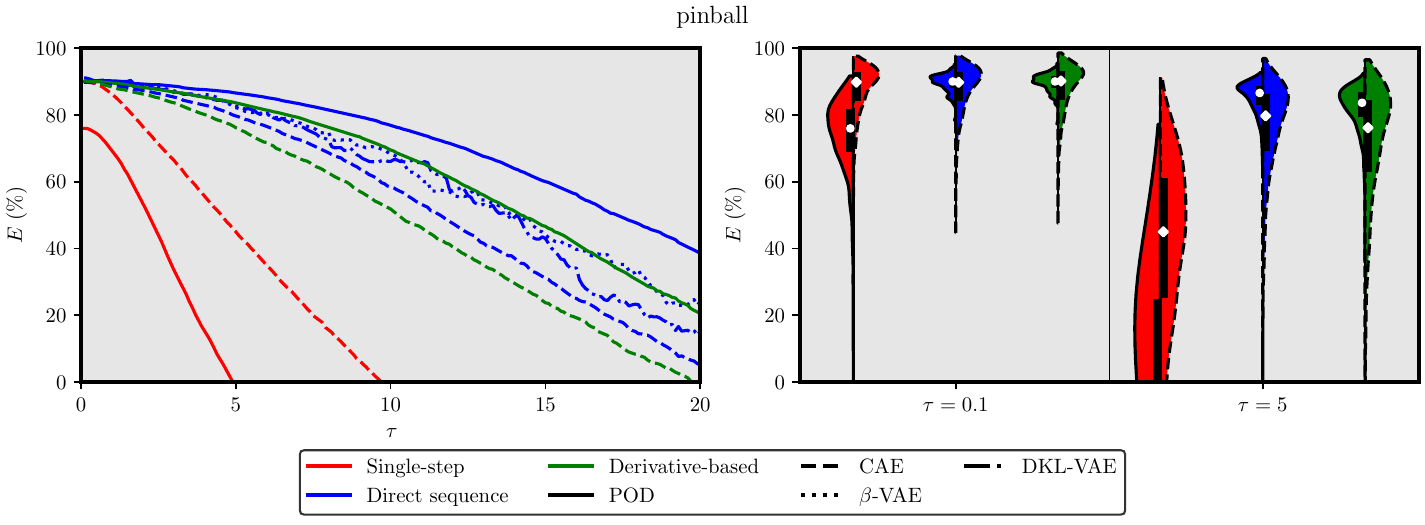}
    \caption{Prediction accuracy $E$ for the fluidic pinball over an ensemble of 20 training seeds. Left: median of $E$ (Eq.~\ref{ec:E}) as a function of the prediction horizon $\tau$ (convective time $t_c$ units), taken over all prediction windows in the test set and all seeds. Colours denote the \ac{LSTM} architectures (\texttt{Single-step}, \texttt{Direct sequence}, \texttt{Derivative-based}); line styles denote the encoder: solid \ac{POD}, dashed \ac{CAE}, dotted \ac{bVAE}, dash-dotted \ac{dklVAE}. Right: distributions of the instantaneous $E$ (evaluated snapshot by snapshot) for the \ac{POD}- and \ac{CAE}-based models, pooling all prediction windows and seeds, at $\tau=0.1$ and $\tau=5$. White markers indicate the median ($\circ$ for \ac{POD}, $\Diamond$ for \ac{CAE}); black boxes span the interquartile range (25th--75th percentiles). The first reported horizon is $\tau=0.1$.}
    \label{fig:Eplot}\label{fig:Epinball}
\end{figure}

\begin{figure}[ht!]
    \centering
    \includegraphics[width=\linewidth]{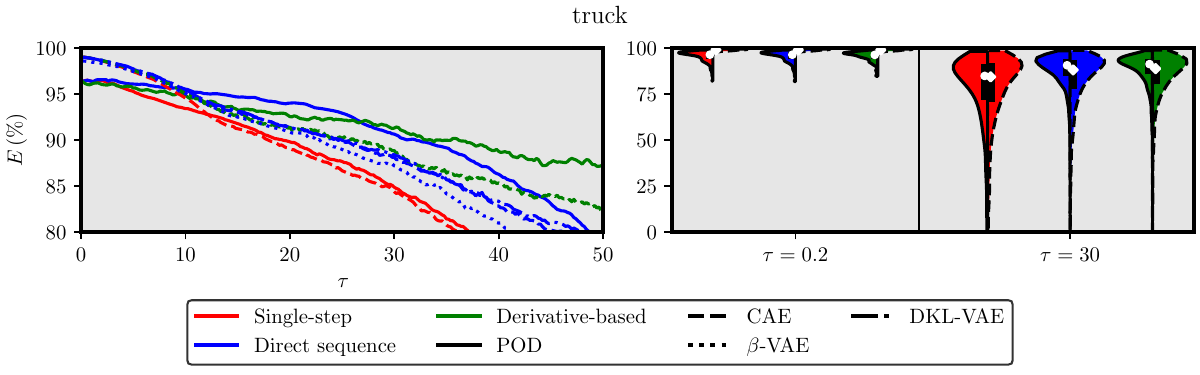}
    \caption{
    Prediction accuracy $E$ for the simplified truck over an ensemble of 20 training seeds, with the same protocol, statistics and line conventions as Figure~\ref{fig:Epinball}. Left: median of $E$ over all prediction windows and seeds as a function of $\tau$, reported up to $\tau=50$ (see text). Right: distributions of the instantaneous $E$ for the \ac{POD}- and \ac{CAE}-based models at $\tau=0.2$ and $\tau=30$. The first reported horizon is $\tau=0.2$.}
    \label{fig:violin}\label{fig:Etruck}
\end{figure}

Having established the compression baseline, we now assess the core task: the long-term prediction of the latent-space dynamics. The left panels of Figures~\ref{fig:Epinball} (fluidic pinball) and~\ref{fig:Etruck} (simplified truck) show the accuracy of the prediction, $E$, as a function of the prediction time horizon, $\tau$. This accuracy metric $E$ is computed on the reconstructed flow fields, thus measuring the performance of the whole framework from end-to-end. To quantify the sensitivity of the comparison to training stochasticity, the complete framework is retrained for an ensemble of 20 independent seeds per case: for each seed, the \ac{CAE}, \ac{bVAE} and \ac{dklVAE} encoders are retrained from scratch and the \ac{LSTM} predictors are trained on the latent space of the same-seed encoder, while the \ac{POD} basis is deterministic and only its predictors are re-seeded. The curves report, at each horizon, the median of $E$ over all prediction windows in the test set and all seeds, and the violin panels pool the corresponding instantaneous values.

A consistent trend emerges across both flow cases. The \texttt{Single-step} models (in red) exhibit a rapid decay in accuracy, as the recursive error accumulation quickly destabilises the prediction. The \texttt{Direct sequence} (blue) and \texttt{Derivative-based} (green) models are more stable, generally retaining high accuracy for longer prediction horizons.

However, the most critical finding is the comparison between the compression methods. For the simplified truck case (Figure~\ref{fig:Etruck}), the \ac{AE}-based models (dashed lines) initially show a slight advantage due to the higher \acp{AE} compression power, but for horizons longer than $\tau \approx 10$, the \ac{POD}-based models (solid lines) are more accurate and stable. The separation between encoders nevertheless remains within a few percentage points of $E$ over the reported truck horizons, indicating that this case lacks the dynamical complexity to discriminate strongly between compression methods; the fluidic pinball provides the more demanding comparison. This trend is even more pronounced in the pinball case (Figure~\ref{fig:Epinball}). Despite the superior reconstruction performance of the \acp{AE}, the \ac{POD}-based models (solid blue and green lines) consistently outperform their \ac{AE}-based counterparts (dashed blue and green lines) for nearly the entire prediction horizon. This identifies a practical crossover: for short horizons, the \acp{AE} can be competitive due to their compactness, whereas for horizons beyond roughly $\tau \approx 10$ (truck) and essentially across all tested horizons (pinball), \ac{POD} yields more reliable forecasts.

The \ac{dklVAE}, in dash-dotted lines in Figures~\ref{fig:Epinball} and~\ref{fig:Etruck}, follows the same pattern as the other nonlinear encoders. For the truck ($d=8$) and the \texttt{Direct sequence} model, its accuracy at the first reported horizon, a median $E\approx99\%$, is indistinguishable from the \ac{CAE}; at intermediate horizons it remains at \ac{CAE} level, with a median $E=91.3\%$ at $\tau=20$ against $91.4\%$ for the \ac{CAE}, $90.9\%$ for the \ac{bVAE} and $94.0\%$ for \ac{POD} (\texttt{Direct sequence} predictors), and at $\tau=50$ it stays below the best \ac{POD}-based model ($78.9\%$ against $87.4\%$).
For the pinball ($d=10$), the \ac{dklVAE} accuracy at the first reported horizon is again in line with the other nonlinear encoders (median $E\approx 91.1\%$, between the \ac{CAE} and \ac{bVAE} values); its forecast accuracy is comparable to the \ac{CAE} and \ac{bVAE} at both the short and long ($\tau=5$, median $E\approx 82.6\%$) horizons used elsewhere in the manuscript, while \ac{POD} remains more accurate throughout. In both cases the more selective regularisation does not recover the long-horizon prediction accuracy of \ac{POD}. The trade-off between compactness and predictability therefore persists when the latent regularisation is decomposed into independently weighted information-theoretic terms.

For the truck, the horizon reported in Figure~\ref{fig:Etruck} is limited to $\tau=50$. Beyond this horizon the run-to-run variability across training seeds grows rapidly: at $\tau=100$ a substantial fraction of the 20 seeds produces diverged predictions (median $E<0$ over the prediction windows) for every encoder--predictor combination except the \ac{POD}-based \texttt{Derivative-based} model, so conclusions drawn at longer horizons would reflect training stochasticity rather than the choice of encoder. The seed analysis also identifies the \texttt{Derivative-based} predictor as the most robust option at long horizons in this case: combined with \ac{POD}, it shows no diverged seeds at $\tau=50$ and retains a median $E=80.5\%$ at $\tau=100$, with a single mild failure across the ensemble.

This result is further clarified by the violin panels of Figures~\ref{fig:Epinball} and~\ref{fig:Etruck}, which show the statistical distribution of the prediction accuracy $E$ of an ensemble of test trajectories and training seeds in both a short- and long-term prediction horizon. At the single-step, short prediction horizon ($\tau = 0.1$ for pinball and $\tau = 0.2$ for truck), all models perform well, with high median accuracies (white dots) and compact distributions. However, at the long horizon (e.g., $\tau = 5$ or $\tau = 30$), the distributions diverge significantly. The \ac{POD}-based models (solid-line violins) maintain a higher median accuracy and generally a narrower distribution, indicating a more reliable performance. In contrast, \ac{CAE}-based models (dashed-line violins) exhibit a lower median accuracy and a very long tail towards $E=0$. This long tail reveals that a large fraction of the \ac{CAE}-based predictions fail significantly, diverging from the true solution. This implies greater sensitivity of the nonlinear latent dynamics to minor prediction errors, so some trajectories drift into qualitatively different regimes and abruptly fail. From a \ac{MPC} standpoint, this means that a slightly less compact but more predictable latent model can enlarge the feasible prediction horizon and, in principle, improve closed-loop robustness. Pooling the training seeds does not alter this picture: the long tail of the \ac{CAE}-based distributions is a property of the models rather than of a particular training realisation.

\subsection{\label{subsec:dynamics_analysis}Analysis of latent-space dynamics and flow prediction}

The contrast between encoders observed in the previous section can be anticipated from the framing of \S~\ref{sec:introduction} and \S~\ref{subsec:predictors}.
 The \ac{POD} basis inherits a structure for shedding-dominated flows analogous to the dynamics observed with a Koopman embedding, whereas the \acp{AE} optimise pointwise reconstruction. It is nonetheless important to distinguish between a linear decomposition and a linear dynamical representation. \ac{POD} provides the former, but not the latter. It is indeed well known that, after projection of the Navier-Stokes equations onto a \ac{POD} basis, the reduced coefficients generally evolve according to nonlinear equations containing modal interaction terms \citep{noack2005need}.

\begin{figure}[ht]
    \centering
    \includegraphics[width=\linewidth]{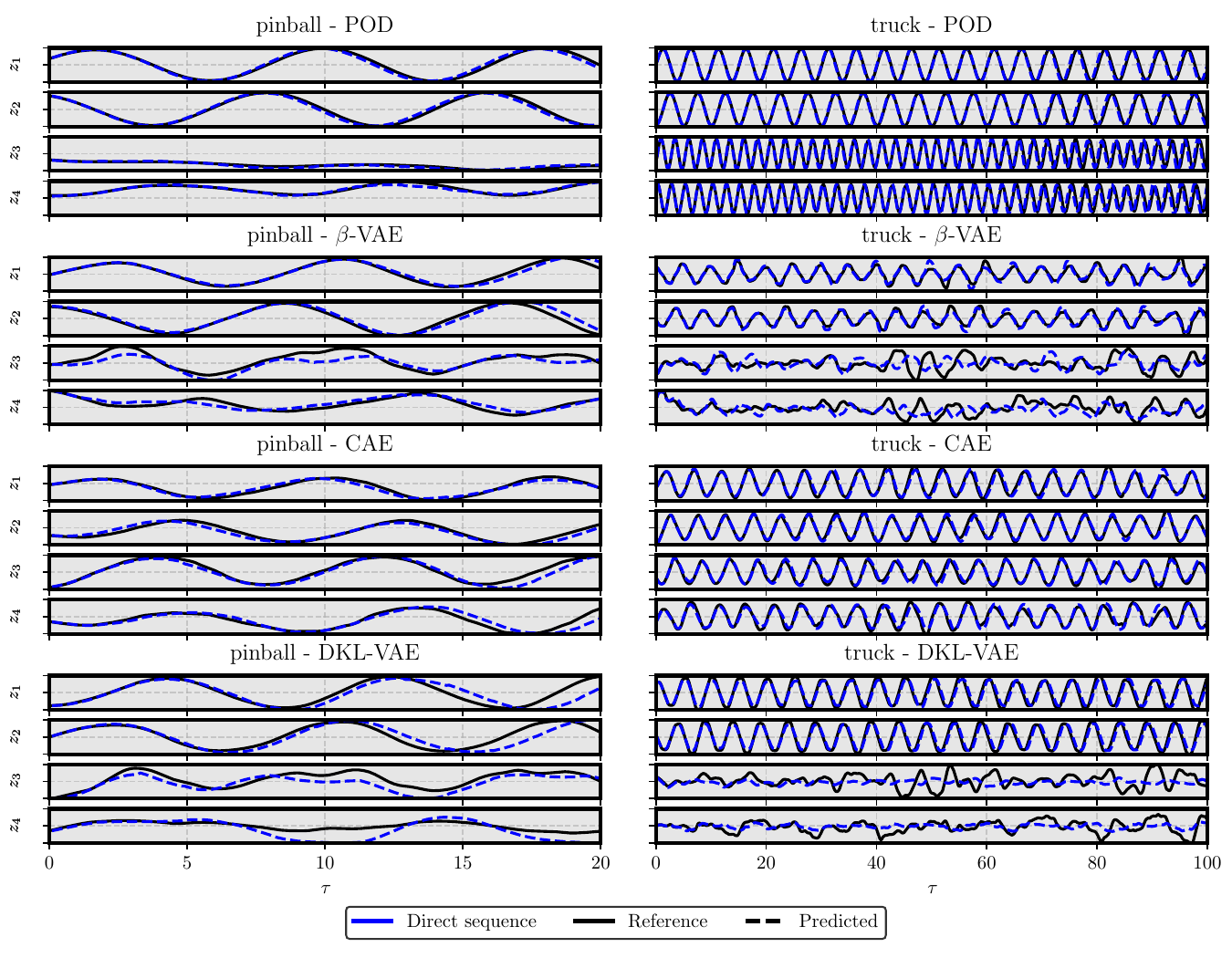}
    \caption{Temporal evolution of the first  four leading latent coefficients (energy-ordered for \ac{POD}; ordered by peak \ac{PSD} of the encoded reference for \ac{CAE}, \ac{bVAE} and \ac{dklVAE}), comparing the ground truth (solid black line) with the prediction of the best-performing \texttt{Direct sequence} \ac{LSTM} model (blue dashed line). Left column shows pinball case and right column truck case.  Rows show \ac{POD} (top), \ac{bVAE} (second), \ac{CAE} (third) and \ac{dklVAE} (bottom) latent spaces.}
    \label{fig:latents}
\end{figure}

The discrepancy in predictive performance, despite the superior encoding capabilities of the \acp{AE}, can be explained by the structure of the latent spaces. Figure~\ref{fig:latents} plots the temporal evolution of the first four latent coefficients, comparing the ground truth ``Reference'' (solid black line) with the prediction from the best-performing \texttt{Direct sequence} \ac{LSTM} model (blue dashed line).

The upper row shows the latent space of the \ac{POD}. The dynamics are simple, smooth, and quasi-periodic, dominated by low-frequency oscillations. As a result, the \ac{LSTM} model can easily learn and accurately forecast this simpler dynamics, with the predicted signal tracking the reference for long durations. 

The remaining rows show the \ac{bVAE}, \ac{CAE} and \ac{dklVAE} latent spaces. In contrast with the \ac{POD}  modes, the dynamics are more complex and non-periodic, showing higher-frequency components. The \acp{AE} encode the flow features into a more compact representation whose dynamics are more irregular and broadband. The \ac{LSTM} has difficulty modelling these more complex dynamics, and the predicted signal diverges from the reference at shorter prediction horizons.

This analysis supports the hypothesis that the linear basis of \ac{POD} creates a latent space with simple, low-frequency dynamics that are easier to learn and predict. Conversely, the nonlinear \acp{AE}, in their less-restricted search for optimal compression, create a more complex and chaotic latent space that is more difficult to forecast.

This conclusion is visualised and further explained by the ``Predicted'' columns of Figure~\ref{fig:fields}. These fields show the final prediction of the flow field at a long-time horizon. They correspond to a single representative trajectory; the quantitative accuracy is reported as the median curves of Figures~\ref{fig:Epinball} and~\ref{fig:Etruck}, where the median of $E$ is taken over $\sim$100 test initial conditions and 20 training seeds.
For the pinball case ($\tau = 20$), the \ac{POD}-based prediction retains the large-scale vortex shedding structure, while showing clear signs of smoothing of the flow fields. The \ac{AE}-based predictions are sharper, retaining smaller details in the flow field but achieving lower accuracy for this snapshot; the \ac{dklVAE} prediction shows the same sharper, lower-accuracy character as the other nonlinear encoders.
For the truck case in $\tau = 50$, all models retain the wake structure, with the \ac{CAE} and \ac{bVAE} predictions slightly below the reconstruction of the \ac{POD} model; the \ac{dklVAE} prediction is consistent with the other nonlinear encoders, following the statistical trend observed in Figures~\ref{fig:Epinball} and~\ref{fig:Etruck}.
It can be noted that the higher spatial frequency content of the \ac{AE} decoded fields, compared to \ac{POD} reconstructed fields, produces better reconstructed energy in the short term predictions, but tends to produce faster error growth as the errors accumulate in the latent-space prediction. This faster error growth can be attributed to the accumulation of phase-shift errors in the predictions, translating into a higher error when the predicted shifted and reference wake flow fields are compared and evaluated (see Figure~\ref{fig:phase_amp_main} below). This behaviour can be identified in Figure~\ref{fig:error}, where the averaged $L_2$ error of the velocity predictions is shown. The plot highlights areas where the predicted fields produce larger errors, coincident with those areas including higher spatial frequency fluctuations. This behaviour is consistent with the idea that nonlinear latent coordinates, while efficient for reconstruction, can exhibit higher sensitivity to initial-condition and phase errors. Hence, small \ac{LSTM} prediction mismatches grow more quickly in the \acp{AE} latent system, effectively reducing its long-horizon predictability. This observation is consistent with the closure perspective of \S~\ref{subsec:predictors}: the broadband \ac{AE} latent dynamics places a heavier burden on the truncated memory term that the predictor must absorb, whereas the narrow-band \ac{POD} dynamics renders the finite-memory approximation closer to sufficient.

\begin{figure}[ht!]
    \centering
    \includegraphics[width=0.8\linewidth]{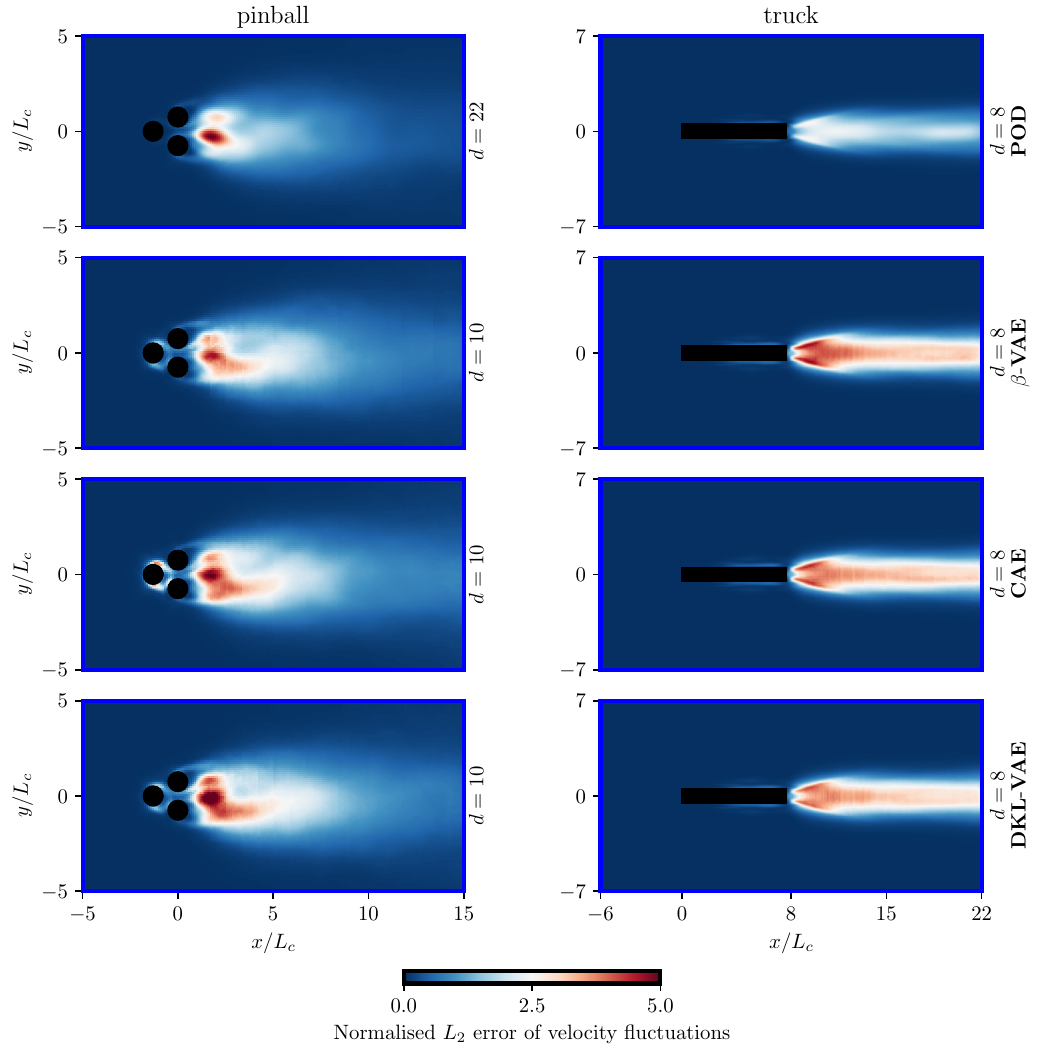}
    \caption{
    Pointwise normalised $L_2$ error magnitude of the predicted velocity fluctuations, combining the streamwise and crosswise components as $\sqrt{e_u^2 + e_v^2}$ at each grid point, where each component $e_i = \langle(\hat{u}_i - u_i)^2\rangle/\sigma_i^2$ is the mean squared error normalised by the variance $\sigma_i^2$ of the corresponding reference velocity component. Each per-component field is averaged over the full prediction horizon, $\tau \in [0, 20]$ for the fluidic pinball (left) and $\tau \in [0, 100]$ for the simplified truck (right), and further averaged over all prediction windows in the test set. Predictions correspond to the best-performing \texttt{Direct sequence} \ac{LSTM} model. Top row: \ac{POD}; second row: \ac{bVAE}; third row: \ac{CAE}; bottom row: \ac{dklVAE}.}
    \label{fig:error}
\end{figure}

\begin{figure}[ht!]
    \centering
    \includegraphics[width=\linewidth]{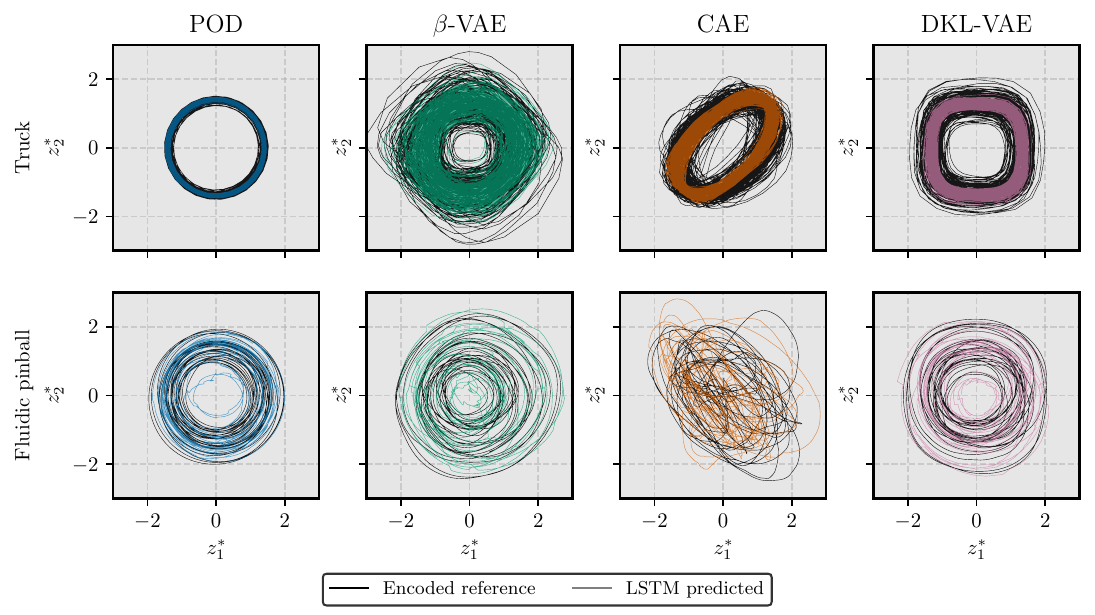}
    \caption{Phase portraits of the dominant standardised latent pair $(z_1^*, z_2^*)$ for the encoded reference (black) and the \texttt{Direct sequence} \ac{LSTM} prediction (coloured). Rows: truck (top) and fluidic pinball (bottom). Columns: \ac{POD}, \ac{bVAE} , \ac{CAE} and \ac{dklVAE}. Standardisation uses the variance of the encoded reference. The \ac{POD} portrait is the most regular for both flows; the \ac{CAE} shows a broader, irregular geometry that the predictor fails to reproduce; the \ac{dklVAE} portrait is comparably structured to the \ac{bVAE}.}
    \label{fig:phase_portraits_main}
\end{figure}

\begin{figure}[ht!]
    \centering
    \includegraphics[width=\linewidth]{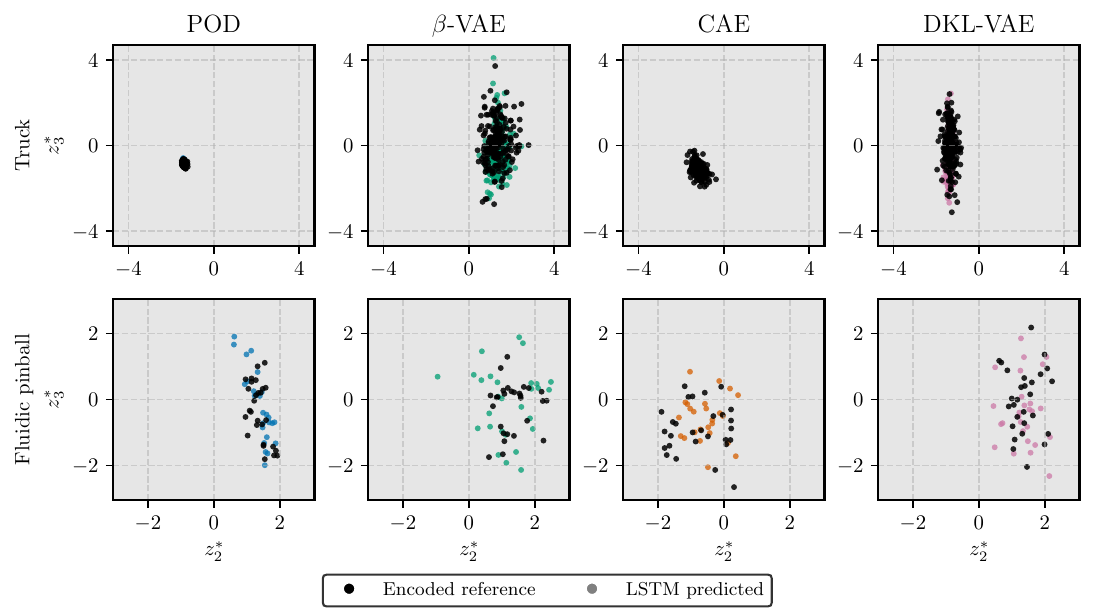}
    \caption{Poincar\'e sections in $(z_2^*, z_3^*)$ taken at positive crossings of $z_1^* = 0$. Encoded reference (black) versus \ac{LSTM}-predicted (coloured), with the same row/column layout as Figure~\ref{fig:phase_portraits_main}. Compact structures indicate quasi-periodic or weakly chaotic behaviour; wider clouds indicate broader spectral content.}
    \label{fig:poincare_main}
\end{figure}

In what follows, \textit{latent-space stability} refers to the predictive accuracy and reliability of the latent dynamics over the admissible control range considered in the dataset. A stable latent representation is one whose multi-time-delay predictor remains bounded, accumulates error slowly over the prediction horizon, and preserves the dominant phase, amplitude and geometric structure of the encoded trajectory. This is an operational, finite-horizon definition tailored to predictive \acp{ROM} for control, not a statement of Lyapunov stability of the full controlled flow. The qualitative differences between the encoder latent spaces can be made quantitative through phase-space diagnostics applied to the best-performing \texttt{Direct sequence} predictor. Figures~\ref{fig:phase_portraits_main} and~\ref{fig:poincare_main} show the phase portraits on the dominant latent pair and a Poincar\'e section in the next two coordinates, comparing the encoded reference with the predicted trajectory. For the truck, the \ac{POD} latent trajectory exhibits a ring-like phase portrait and a compact Poincar\'e section, both well reproduced by the \ac{LSTM}. The \ac{CAE} trajectory shows a broader, irregular geometry, and the predicted Poincar\'e section is markedly more concentrated than the reference, indicating that the \ac{CAE} predictor collapses onto a narrower effective attractor rather than only accumulating phase errors. The \ac{bVAE} sits between the two, with a more regular phase portrait due to the latent space KL regularisation, but a still-broad Poincar\'e section. The \ac{dklVAE} shows a phase portrait and Poincar\'e section comparably structured to the \ac{bVAE}, consistent with the comparable degree of latent regularisation applied by both objectives. The fluidic pinball case shows analogous behaviour with the wider attractor expected from its weakly chaotic dynamics.

To separate the contribution of phase from amplitude in the prediction error, the dominant standardised latent pair is mapped to its polar form, $A(t) = \sqrt{z_1^2 + z_2^2}$ and $\phi(t) = \arg(z_1 + i z_2)$ on the continuous branch. For \ac{POD} the coordinates are energy-ordered; for \ac{bVAE}, \ac{CAE} and \ac{dklVAE} the coordinates are reordered by decreasing peak intensity of the \ac{PSD} of the encoded reference, so that the dominant oscillatory pair is identified consistently across encoders. The phase error is wrapped to $[-\pi,\pi]$, so $|\Delta\phi|$ is bounded in $[0,\pi]$ and $|\Delta\phi|=\pi$ indicates complete loss of phase coherence between the predicted and the encoded dominant pair. Figure~\ref{fig:phase_amp_main} reports the absolute phase error $|\Delta\phi|$ and amplitude error $|\Delta A|$ for both flows. In the truck case, all three baseline encoders retain $|\Delta\phi| < \pi/2$ up to $\tau \approx 400$; \ac{bVAE} then saturates close to $\pi$ around $\tau \approx 550$ and \ac{CAE} around $\tau \approx 900$, while \ac{POD} remains below $\pi/2$ over the entire test trajectory ($\tau \approx 1000$). The \ac{dklVAE} loses phase coherence earlier, with $|\Delta\phi|$ exceeding $\pi/2$ from $\tau \approx 200$ onward. For the fluidic pinball $|\Delta\phi|$ grows faster and the four encoders are closer to each other, with \ac{CAE} reaching $|\Delta\phi|\approx\pi$ around $\tau \approx 80$ and \ac{POD} and \ac{bVAE} oscillating near $\pi$ from $\tau \approx 150$ onward. The amplitude error stays in a comparable range across encoders, with the only clear separation appearing in the truck \ac{POD} case where it is essentially negligible. The decomposition shows that long-horizon degradation in the latent space is driven by phase drift rather than amplitude distortion, and that for the truck \ac{CAE} and \ac{bVAE} lose phase coherence earlier than \ac{POD}. For the more chaotic pinball, the separation between encoders is less clear, as expected.

\begin{figure}[ht!]
    \centering
    \includegraphics[width=\linewidth]{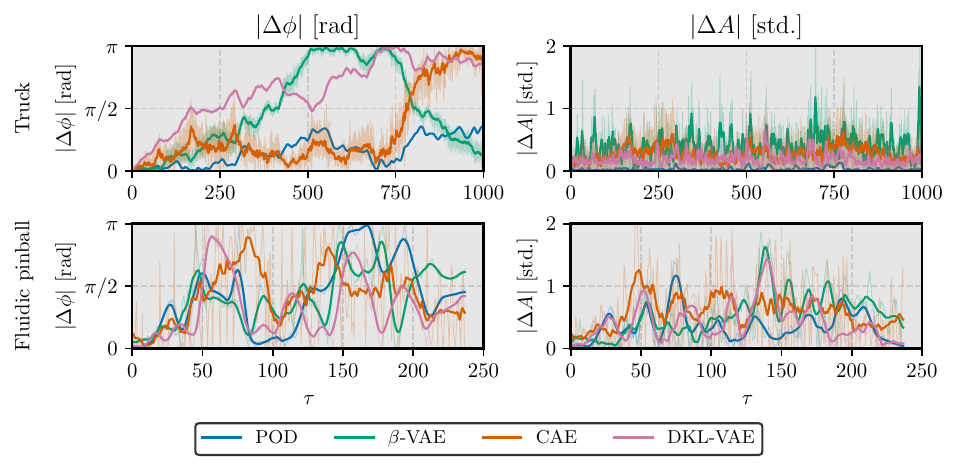}
    \caption{Absolute phase error $|\Delta\phi|$ (left) and absolute amplitude error $|\Delta A|$ (right) of the dominant latent pair, for the truck (top) and fluidic pinball (bottom). Lines: \ac{POD} (blue), \ac{bVAE} (green), \ac{CAE} (orange) and \ac{dklVAE} (pink). Thick lines are a centred moving average over one reference shedding period; thin lines are the raw signals. The errors are similar across encoders except for the truck \ac{POD} case, where the phase error grows more slowly and the amplitude error remains close to zero, indicating that it is dominated by phase error.}
    \label{fig:phase_amp_main}
\end{figure}

\begin{figure}[ht!]
    \centering
    \includegraphics[width=\linewidth]{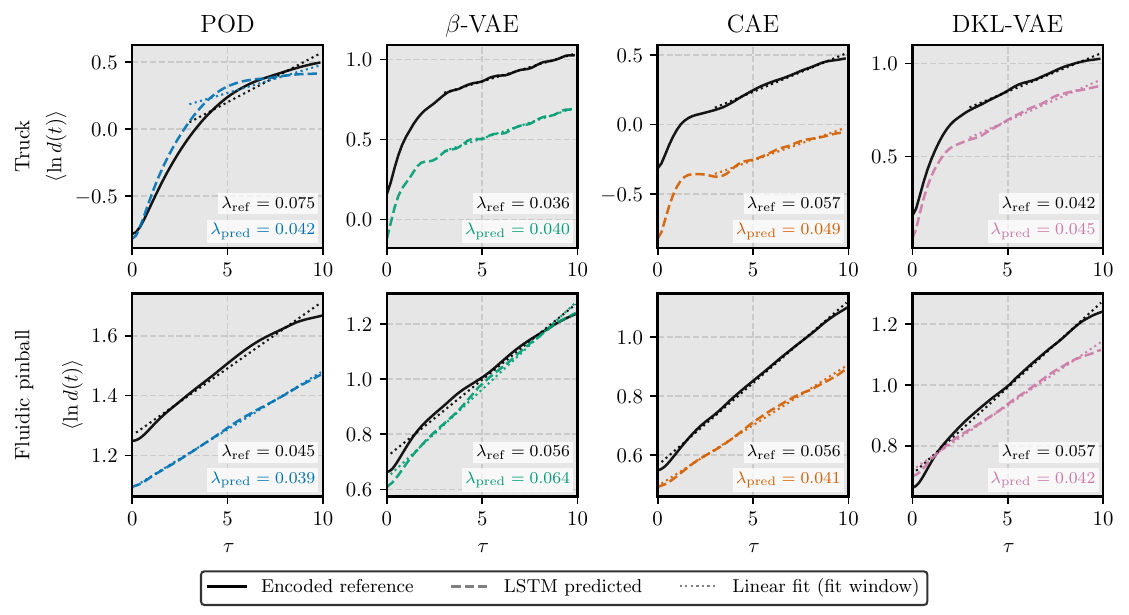}
    \caption{Rosenstein-style logarithmic separation of initially close latent-state pairs, $\langle \ln d(t) \rangle$, for the encoded reference (solid) and the \ac{LSTM}-predicted trajectory (dashed). Rows: truck and fluidic pinball cases. Columns: \ac{POD}, \ac{bVAE} , \ac{CAE} and \ac{dklVAE}. The dotted black line in each panel is the linear fit used to extract the effective short-horizon divergence rate, drawn only within its fit window ($[3, 10]\, t_c$ for the truck, $[0.2, 10]\, t_c$ for the pinball). Fitted slopes are annotated in the top-left corner of each panel as $\lambda_{\mathrm{ref}}$ and $\lambda_{\mathrm{pred}}$. Neither case exhibits a clean asymptotic log-linear regime, reflecting the quasi-periodic or weakly chaotic nature of the actuated wakes rather than a defect of the estimator; $\lambda_{\mathrm{eff}}$ is therefore reported as an integrated short-horizon divergence rate over the fit window, not as a Lyapunov exponent.}
    \label{fig:divergence_main}
\end{figure}

Beyond the phase and amplitude decomposition, we further characterise the latent dynamics through their short-horizon sensitivity to perturbations. We estimate the effective divergence rate from the latent trajectories ~\citep{rosenstein1993practical}. For each reference point $\boldsymbol{z}_i$ in the standardised $d$-dimensional latent state, the nearest neighbour $\boldsymbol{z}_j$ is identified subject to a temporal exclusion window of at least one reference shedding period (estimated from the dominant Welch frequency of the encoded $z_1$), which prevents false neighbours arising from temporal correlations. The control sequence is included in the neighbour-search space so that neighbours share a similar actuation history, while future control inputs are not explicitly considered and its effect is therefore included in the divergence rates. The mean logarithmic separation $\langle \ln d(k\Delta t) \rangle$ is then fitted linearly on a fixed convective-time window, with $[0.2, 10]\, t_c$ for the pinball and $[3, 10]\, t_c$ for the truck. The truck window starts at $3\, t_c$ to skip the initial transient observed in the curves of Figure~\ref{fig:divergence_main}.

We report the resulting slope as an effective short-horizon divergence rate $\lambda_{\mathrm{eff}}$ rather than a Lyapunov exponent. The two flows studied here are not classically chaotic in the Rosenstein sense: the truck wake at $Re = 500$ is quasi-periodic, and the fluidic pinball at $Re = 150$ is only weakly chaotic, just past the transition to chaotic dynamics~\citep{deng2020}. As a consequence, an asymptotic log-linear divergence regime is not observed over the test-trajectory horizon; the mean separation saturates relatively quickly as it approaches a fraction of the attractor size. This is a physical feature of the actuated wakes, not a defect of the estimator. We therefore interpret $\lambda_{\mathrm{eff}}$ as an integrated divergence rate over the fit window rather than as an asymptotic dynamical invariant, and we rely on the relative ordering of the values across methods and the comparison between encoded-reference and predicted trajectories, rather than on the absolute magnitude. Figure~\ref{fig:divergence_main} makes this regime visible: the linear fits are drawn only within the fit window, so one can judge the local quality of the linear approximation for each curve. Under control the latent trajectory is non-autonomous, so the separation of initially close states reflects both the intrinsic sensitivity of the flow and differences in the forcing histories, a contribution that classical Lyapunov exponents do not isolate and that is instead the object of conditional Lyapunov exponents in driven systems~\citep{pecora1990synchronization}; the inclusion of the control sequence in the nearest-neighbour search described above ensures that compared trajectories share similar actuation histories. Indicators of this kind have recently been shown to be recoverable directly in a latent space, although in the autonomous setting~\citep{ozalp2024stability}.

The fluidic pinball results show encoded-reference divergence rates of comparable magnitude across all four encoders ($\lambda_{\mathrm{ref}} \approx 0.045$--$0.057$ per convective time), consistent with the separation rate of the chaotic wake being largely independent of the choice of encoder. The \ac{LSTM} predictors reproduce the same order of magnitude; the \ac{POD} predictor slightly underestimates the reference, the \ac{bVAE} predictor slightly overestimates it, and the \ac{CAE} and \ac{dklVAE} predictors moderately
underestimate it. For the truck, the encoded \ac{POD} does not show a proper log-linear region, consistent with the quasi-periodic trajectory in the \ac{POD} latent space. The value of $\lambda_{\mathrm{ref}}$ annotated for this case should therefore not be compared quantitatively with the other encoders, as it mainly reflects the curvature of the separation curve within the fit window. The \ac{dklVAE} shows a divergence rate comparable to the \ac{CAE} and \ac{bVAE} in both cases, consistent with its similar long-horizon forecast accuracy.

From a practical standpoint, these diagnostics suggest valuable indications to construct a stable latent space for controlled flows. Encoders producing smooth, low-dimensional latent dynamics associated with the dominant coherent oscillatory structures, here \ac{POD}, yield more predictable and therefore more stable latent trajectories than purely reconstruction-driven nonlinear encoders. In addition, standardisation of the latent coordinates, inclusion of the control history in the predictor and in the stability diagnostic, and selection of the latent dimension based on long-horizon predictability rather than reconstruction energy alone improve the conditioning of the latent predictor. For flows undergoing strong control-induced bifurcations, a single global latent space may not be sufficient; in such cases, control-conditioned encoders, local latent models for different control regimes, or regime-aware mixture models may be more appropriate. Imposing stability by construction, for instance through Lyapunov-based reduced-order models for predictive control \citep{zhao2022machine}, is a promising direction for future work.

Overall, the results indicate that for reliable, long-term forecasting, the stability and simplicity of the latent dynamics can be more relevant than the compression efficiency of the encoder, as long as the large-scale structures provide enough information for the application.

\section{\label{sec:conclusion}Conclusions}

We have discussed the effectiveness of data-driven frameworks based on latent-space compression and multi-time-delay embedding for prediction of wake flows under control actions. We compared the classical linear \ac{POD} encoder against a nonlinear \ac{CAE} and two regularised variational variants, a \ac{bVAE} and a \ac{dklVAE}, and assessed three \ac{LSTM}-based predictors, including an autoregressive single-step model, a direct sequence-to-sequence model, and a derivative-based model with explicit time integration. 

As expected, \acp{AE} provided clearly higher compression efficiency than \ac{POD}, achieving comparable reconstruction accuracy with fewer latent variables. Surprisingly, this advantage did not translate into superior long-horizon prediction. For short horizons, \ac{AE}-based models can match or slightly outperform \ac{POD} due to their stronger reconstruction of small-scale features. Yet, as the horizon increases, \ac{POD}-based models become consistently more accurate and markedly more stable. This crossover occurs around $\tau \approx 10$ in the truck case and is essentially present throughout the tested horizons in the more complex pinball dynamics. The violin-plot statistics confirmed that \ac{CAE}-based predictors are more prone to catastrophic divergence at long horizons, whereas \ac{POD}-based predictors retain higher median accuracy and a narrower spread, indicating greater reliability. These statistics are computed over ensembles of 20 independently retrained encoder--predictor pairs per case, so the reported comparison is robust to training stochasticity; for the truck, the same analysis shows that horizons beyond $\tau=50$ are dominated by run-to-run variability, and the reported horizon is clipped accordingly.

We explain this counterintuitive result with the different complexity of the latent space. \ac{POD} yields smooth, quasi-periodic latent trajectories that are easy for \acp{LSTM} to learn and extrapolate. In contrast, \ac{AE} latent variables exhibit more irregular, broadband dynamics. Even when the decoded fields retain sharper spatial detail, the increased nonlinearity and sensitivity of the latent evolution lead to faster accumulation of phase errors, shortening the reliable prediction window. Therefore, in the context of long-horizon forecasting for receding-horizon control, the simplicity and stability of the latent dynamics can be more valuable than maximal compression.

Within the linearisation viewpoint, the long-horizon predictability of \ac{POD} is due to narrow-band dynamics resembling coordinates associated with dominant Koopman spectral components of shedding-dominated wake flows. Such alignment does not generally occur in a reconstruction-trained \ac{AE}. The \ac{bVAE} results do not recover the long-horizon accuracy of \ac{POD} either; for the truck it in fact trails the \ac{CAE} at every horizon beyond $\tau \approx 10$ (Figure~\ref{fig:Etruck}), indicating that latent disentanglement alone is not sufficient to recover this property. The same holds for the \ac{dklVAE} evaluated on the truck and pinball cases: decomposing the latent regularisation into independently weighted information-theoretic terms matches the reconstruction and has a comparable degree of linear independence between latent coordinates of the \ac{bVAE} but does not restore the long-horizon prediction accuracy of \ac{POD}, suggesting that the trade-off is not an artefact of the particular \ac{bVAE} penalty. The trade-off is similarly robust to encoder architecture choices: supplementary experiments on the fluidic pinball, varying the filter count and activation function of the \ac{CAE} over ensembles of training seeds (Appendix~\ref{subsec:annex_architecture}), show no notable effect on long-term predictability, with the baseline \ac{CAE} remaining a well-balanced operating point.

Regarding temporal predictors, we assessed architectures with multi-time-delay embedding of the control vector within the latent space. Even under equal lookback horizons, we observed a significant sensitivity in the prediction accuracy on the choice of the \ac{LSTM} model. Autoregressive single-step models degraded rapidly due to recursive error accumulation. The direct sequence and derivative-based architectures improved stability and yielded the best long-term performance, with the direct sequence model being especially aligned with \ac{MPC} practice because it maps a candidate future control sequence directly into a predicted state trajectory.

Overall, the results highlight a practical principle for data-driven model-based \ac{AFC}: a slightly higher-dimensional but dynamically simpler latent representation may enable more reliable prediction over longer horizons. 
An essential practical caveat is that the relative performance of \ac{AE}- versus \ac{POD}-based \acp{ROM} also depends on the capacity of the temporal predictor. In principle, more expressive sequence models could better accommodate the broadband \ac{AE} latent dynamics, shifting the trade-off in favour of nonlinear encoders. However, such gains must be weighed against real-time compute, latency, and power constraints for deployment on embedded platforms, motivating co-design of the encoder and predictor with hardware feasibility in mind. Furthermore, the validation in this work relied on simulation data at relatively low Reynolds numbers, characterised by the absence of noise and compact representability using \ac{POD}. It can be hypothesised that,  for flows exhibiting a broader bandwidth of energetically relevant frequencies in their velocity fluctuation spectra, linear techniques would eventually be superseded by nonlinear compression methods based on \acp{AE}. Future work will be targeted to assess this aspect at higher Reynolds numbers and in different flow configurations.

The results presented in this work should be interpreted within the limitations of the chosen flow configurations. The two-dimensional, low-Reynolds-number regime does not capture the full three-dimensional complexity, broadband spectral content or chaotic dynamics of fully-developed turbulence. Future developments should also focus on testing actuation-aware or regularised nonlinear encoders that explicitly trade spatial fidelity for latent predictability, assess these ideas in three-dimensional and higher-Reynolds-number settings, and close the loop within real-time \ac{MPC} experiments to quantify the control benefit of longer stable horizons.

\section*{Declaration of competing interest}
The authors report no conflict of interest.

\section*{Data availability}

All datasets and codes used in this work will be made openly available in public repositories upon publication.

\section*{CRediT authorship contribution statement}

\textbf{Alberto Solera-Rico:} Conceptualization, Methodology, Software, Validation, Formal analysis, Investigation, Data Curation, Writing – Original draft, Visualization.
\textbf{Patricia García-Caspueñas:} Conceptualization, Methodology, Software, Validation, Formal analysis, Investigation, Data Curation, Writing – Original draft, Visualization.
\textbf{Carlos Sanmiguel Vila:} Conceptualization, Methodology, Resources,  Writing – Original draft, Writing – Review \& Editing, Supervision.
\textbf{Stefano Discetti:} Conceptualization, Methodology, Resources, Writing – Review \& Editing, Supervision, Project administration, Funding acquisition.

\section*{Declaration of generative AI in scientific writing}

During the preparation of this work, the authors used Gemini, Claude and ChatGPT to improve the readability and language of the manuscript. After using these tools, the authors reviewed and edited the content as needed and assume full responsibility for the content of the published article.

\section*{Funding sources}

This project has received funding from the European Research Council (ERC) under the European Union’s Horizon 2020 research and innovation programme (grant agreement No 949085, NEXTFLOW ERC StG). Views and opinions expressed are however those of the authors only and do not necessarily reflect those of the European Union or the European Research Council. Neither the European Union nor the granting authority can be held responsible for them. 


\appendix

\section{\label{sec:annex_sensitivity}Sensitivity analyses}

This appendix reports the sensitivity studies that support the conclusions of \S~\ref{sec:results}, covering the sampling frequency of the latent-space time series, the capacity and architecture of the encoders, and an alternative latent-space predictor based on \ac{SINDy}~\citep{brunton2016sindy}. All studies use the fluidic pinball case unless stated otherwise, with the \texttt{Direct sequence} \ac{LSTM} predictor as the reference dynamical model.

\subsection{Sampling frequency}

\begin{figure}[ht!]
    \centering
    \includegraphics[width=\linewidth]{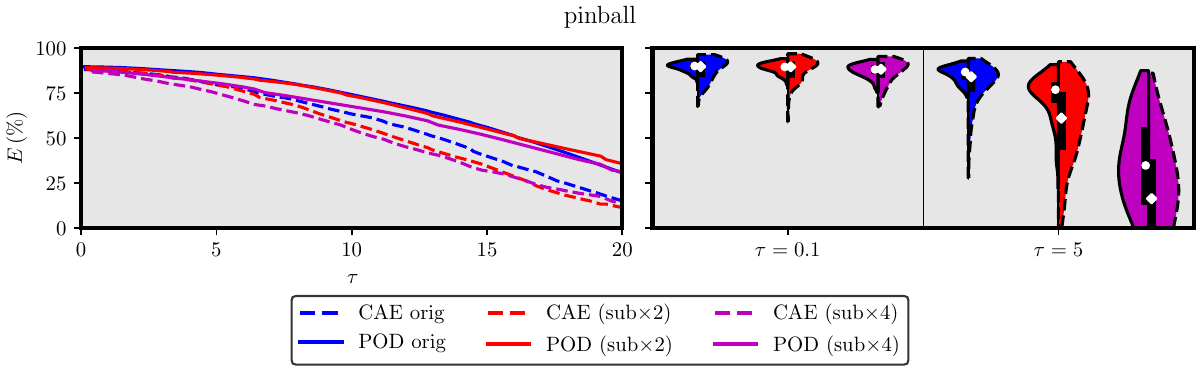}
    \caption{Effect of the sampling rate of the latent-space time series on the prediction performance for the fluidic pinball with $E=90\%$ and the \texttt{Direct sequence} \ac{LSTM} predictor. Three rates are compared: $\Delta t$, $2\Delta t$ and $4\Delta t$. (Left) Prediction accuracy $E$ as a function of the prediction horizon $\tau$. (Right) Violin plots of $E$ at a short-term and a long-term horizon. White markers indicate the median ($\circ$ for \ac{POD}, $\Diamond$ for \ac{CAE}); black boxes span the interquartile range (25th--75th percentiles). }
    \label{fig:annex_undersampled}
\end{figure}

\begin{figure}[ht!]
  \centering
  \includegraphics[width=\linewidth]{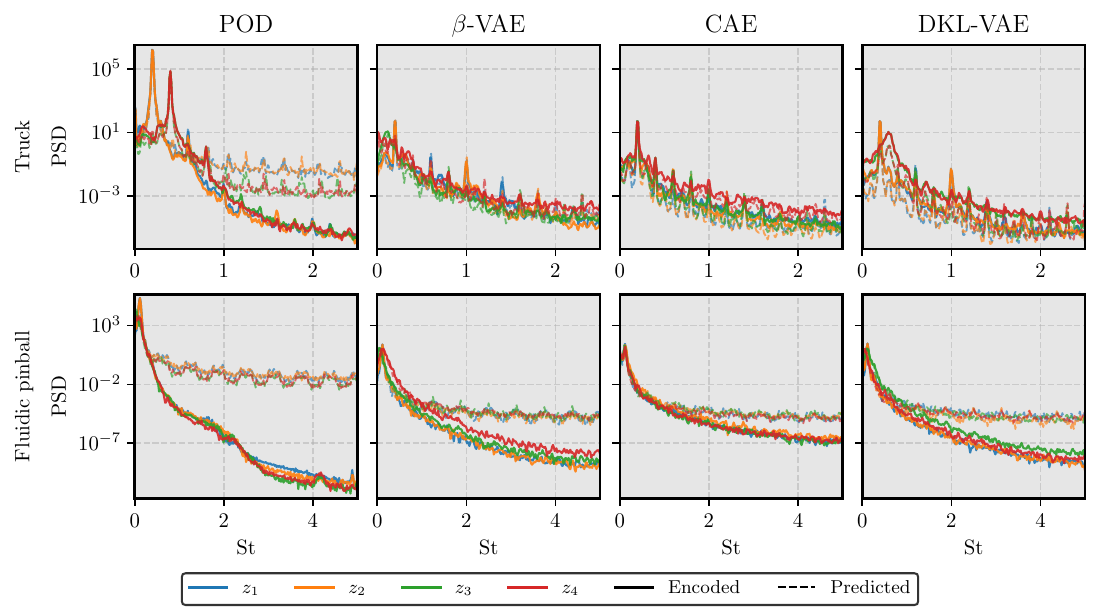}
  \caption{Power spectral density against Strouhal number $\textrm{St} = f L_{ref} / U_{\infty}$ of the four leading latent coordinates ($z_1$--$z_4$) for the encoded reference (solid) and the \texttt{Direct sequence} \ac{LSTM} prediction (dashed), with $L_{ref}$ the truck width or cylinder diameter and $U_{\infty}$ the reference velocity. The upper limit of each panel corresponds to the Nyquist frequency of the original sampling rate. Rows: truck (top) and fluidic pinball (bottom). Columns: \ac{POD}, \ac{bVAE} , \ac{CAE} and \ac{dklVAE}. The dominant spectral peaks are well resolved at the original sampling rate for all encoders, with several orders of magnitude separating the peaks from the values at the Nyquist frequency.}
  \label{fig:psd}
\end{figure}

The original sampling interval was selected to resolve the dominant frequency content of the latent variables while keeping memory and computational cost within practical bounds. To assess the sensitivity of the prediction performance to this choice, we repeated the analysis with two coarser sampling rates, $\Delta t = 0.2$ and $\Delta t = 0.4$ ($2\times$ and $4\times$ the original step). The results in Figure~\ref{fig:annex_undersampled} at $2\Delta t$ are very close to those at the original $\Delta t$, confirming that the original sampling rate is sufficient to capture the relevant latent dynamics. The \ac{CAE} is slightly more affected by undersampling than \ac{POD}, consistent with the broader spectral content of its latent trajectories, but the gap is not large enough to explain alone the performance difference observed at the original $\Delta t$. The relative ranking of \ac{POD} and \ac{CAE} is preserved across all tested rates. The \ac{PSD} of the latent coordinates, reported in Figure~\ref{fig:psd}, further confirms that the dominant frequencies are well resolved for both encoders, with several orders of magnitude separating the spectral peaks from the values at the Nyquist frequency of each sampling rate.

\subsection{Encoder capacity}

\begin{figure}[ht!]
    \centering
    \includegraphics[width=\linewidth]{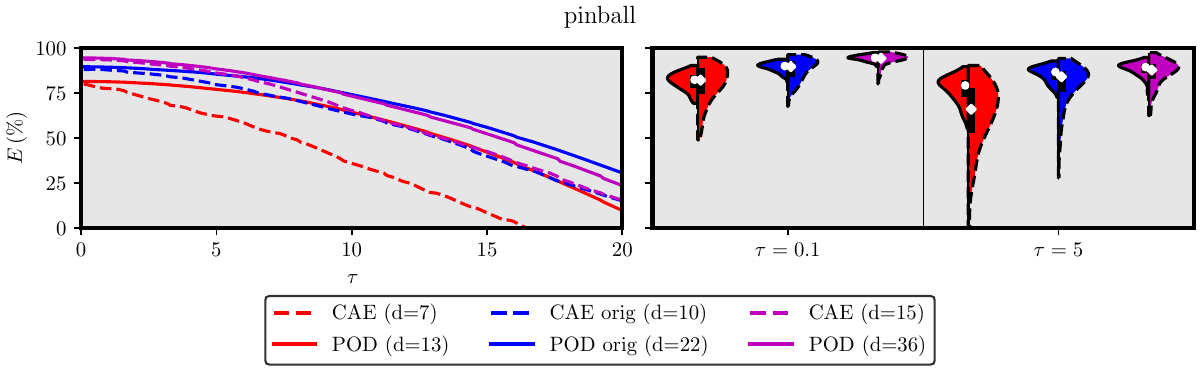}
    \caption{Effect of the encoder capacity on the prediction performance for the fluidic pinball trained with the \texttt{Direct sequence} \ac{LSTM} predictor, for three reconstructed energy targets: $E=80\%$, $90\%$ (baseline) and $95\%$. (Left) Prediction accuracy $E$ as a function of the prediction horizon $\tau$. (Right) Violin plots of $E$ at a short-term and a long-term horizon. White markers indicate the median ($\circ$ for \ac{POD}, $\Diamond$ for \ac{CAE}); black boxes span the interquartile range (25th--75th percentiles).}
    \label{fig:annex_latent}
\end{figure}

To rule out underfitting as the source of the lower forecast accuracy of the \ac{CAE}, we performed a latent-dimension sweep for both encoders, selecting $d$ values corresponding to $E = 80\%$, $90\%$ (baseline), and $95\%$ of the reconstructed variance. This yields \ac{CAE} models with $d = 7,10,15$ and \ac{POD} models with $d = 13,22,36$, see Figure~\ref{fig:annex_latent}. Increasing the reconstructed energy reduces the prediction variance for short-to-medium horizons (up to approximately $\tau = 5$). At long horizons, however, the baseline $E = 90\%$ remains the most reliable configuration for both encoders. Larger latent dimensions improve short-term accuracy but degrade long-term stability, confirming that the compactness--predictability trade-off is a fundamental property of the framework rather than a consequence of underfitting at the baseline configuration. The relative ranking of \ac{POD} and \ac{CAE} is preserved across all tested dimensions. We also note that the capacity of the baseline \ac{CAE} model was previously optimised through a hyperparameter study.

\subsection{\texorpdfstring{Encoder architecture}{Encoder architecture}}
\label{subsec:annex_architecture}

Beyond the latent dimension, the encoder architecture itself is a potential source of bias in the compactness--predictability comparison. To assess its influence, we varied the number of convolutional filters and the activation function of the \ac{CAE} encoder--decoder ($d=10$, all other hyperparameters fixed). The notation CAE$_N$ denotes a variant whose convolutional layers use $N$ times the number of filters of the baseline CAE$_1$; CAE$_{\mathrm{tanh}}$ uses a hyperbolic tangent activation in place of the baseline ELU, with the same filter count as CAE$_1$. For each encoder the same \texttt{Direct sequence} \ac{LSTM} predictor is trained with identical settings, and each variant is independently retrained for 20 seeds, following the protocol of \S~\ref{sec:results}.

\begin{figure}[ht!]
    \centering
    \includegraphics[width=\linewidth]{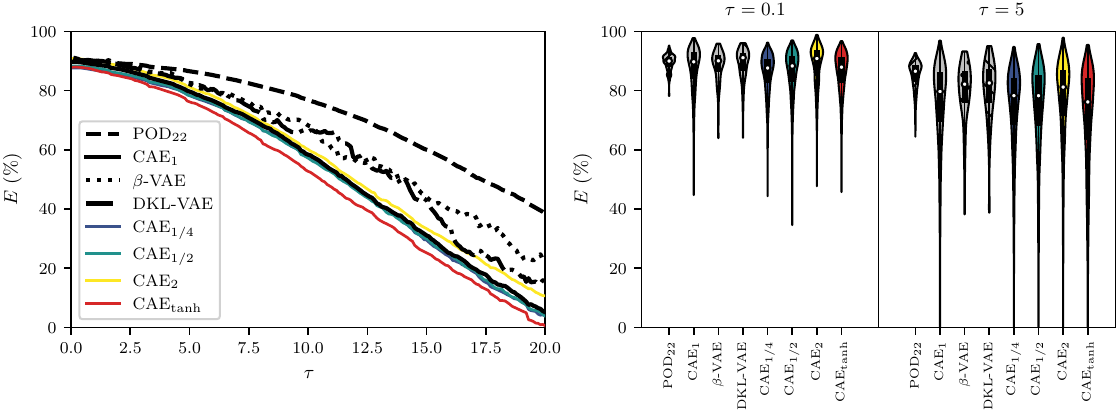}
    \caption{Median prediction accuracy $E(\tau)$ over the training-seed ensemble (left) and pooled distributions of the instantaneous $E$ at $\tau=0.1$ and $\tau=5$ (right) for the paper baselines (POD$_{22}$, CAE$_1$, \ac{bVAE}, \ac{dklVAE}) and the architecture variants, evaluated on the fluidic pinball case with $d=10$ and the \texttt{Direct sequence} \ac{LSTM}. Numerical subscripts denote the filter-count multiplier relative to CAE$_1$; CAE$_{\mathrm{tanh}}$ has the same filter count as CAE$_1$ but uses a hyperbolic tangent activation.}
    \label{fig:annex_architecture}
\end{figure}

The results are reported in Figure~\ref{fig:annex_architecture}. Across the tested range, the architecture has only a marginal effect on long-term predictability: CAE$_2$ performs marginally better than the baseline (median $E=81.2\%$ against $79.7\%$ for CAE$_1$ at $\tau=5$), while CAE$_{1/2}$, CAE$_{1/4}$ and CAE$_{\mathrm{tanh}}$ remain within $3.5$ percentage points of the baseline, and no variant approaches the long-horizon accuracy of \ac{POD} (median $E=86.6\%$ at $\tau=5$). The relative ranking of \ac{POD} and the \ac{CAE} variants is preserved, and the baseline CAE$_1$ remains a well-balanced operating point between capacity and accuracy. These results indicate that the compactness--predictability trade-off is a robust property of the encoder family and is not an artefact of the specific architecture chosen for the baseline comparison.

\section{SINDy as latent-space predictor}

\begin{figure}[ht!]
    \centering
    \includegraphics[width=\linewidth]{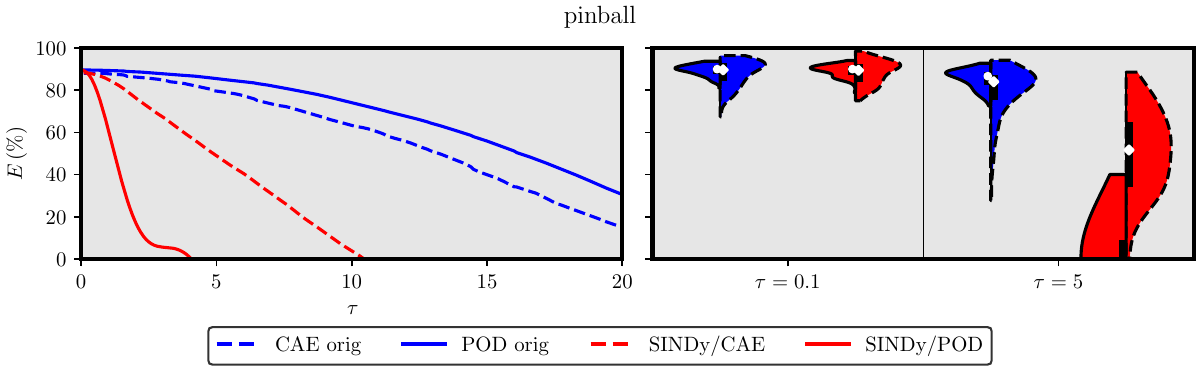}
    \caption{Comparison between \ac{SINDy} and the \texttt{Direct sequence} \ac{LSTM} predictor on the fluidic pinball with $E=90\%$. (Left) Prediction accuracy $E$ as a function of the prediction horizon $\tau$. (Right) Violin plots of $E$ at a short-term and a long-term horizon. White markers indicate the median ($\circ$ for \ac{POD}, $\Diamond$ for \ac{CAE}); black boxes span the interquartile range (25th--75th percentiles).}
    \label{fig:annex_SINDy_pinball}
\end{figure}

\begin{figure}[ht!]
    \centering
    \includegraphics[width=\linewidth]{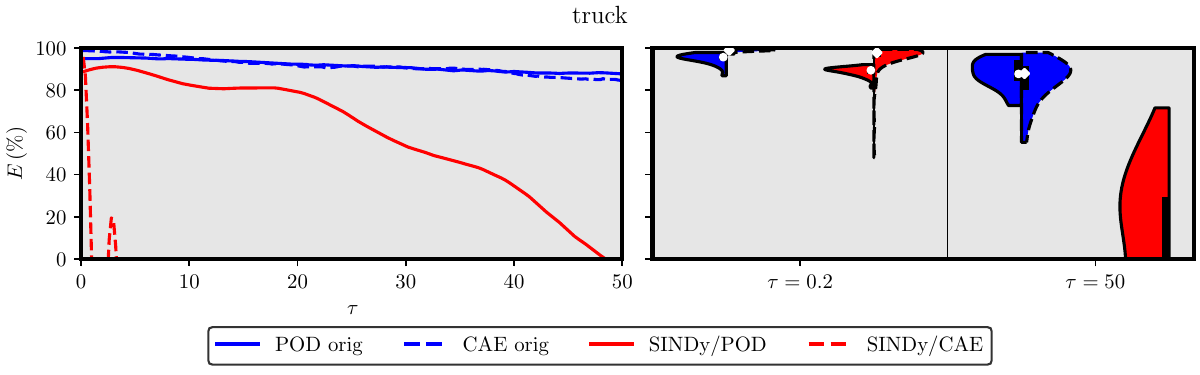}
    \caption{Comparison between \ac{SINDy} and the \texttt{Direct sequence} \ac{LSTM} predictor on the truck case. (Left) Prediction accuracy $E$ as a function of the prediction horizon $\tau$. (Right) Violin plots of $E$ at a short-term and a long-term horizon. White markers indicate the median ($\circ$ for \ac{POD}, $\Diamond$ for \ac{CAE}); black boxes span the interquartile range (25th--75th percentiles).}
    \label{fig:annex_SINDy_truck}
\end{figure}

We finally tested \ac{SINDy} with control inputs \citep{brunton2016sparse} as an alternative to the \ac{LSTM} predictor on both the \ac{POD} and \ac{CAE} latent spaces of the two flow configurations. The regularisation variant used was SINDy-ALASSO, reported by \cite{fukami2021sparse} as a robust sparse identification method for complex dynamics of higher dimensionality. Forward integration of the identified sparse polynomial models without an explicit closure causes the predicted trajectories to diverge from the reference within a few convective times. For the chaotic fluidic pinball, the absence of a closure term makes the identified equations unstable for long-horizon forecasting (Figure~\ref{fig:annex_SINDy_pinball}). The \ac{POD}--\ac{SINDy} model, with $22$ coupled ordinary differential equations, was particularly expensive to identify and still diverged within a few convective times; the \ac{CAE}--\ac{SINDy} model showed a comparable instability. The truck case, with a smaller latent dimension $d=8$, should in principle be more favourable for \ac{SINDy}; nevertheless, the resulting predictors decay faster than the \ac{LSTM} alternatives, Figure~\ref{fig:annex_SINDy_truck}. The approach completely fails to capture the dynamics of the \ac{CAE} latent space and lacks long-term stability for the \ac{POD} modes. Rather than acting as an effective regulariser, the structural sparsity of \ac{SINDy} proves insufficient to compensate for the complexity of the controlled-wake latent dynamics under the conditions tested here. The \ac{LSTM}, with its recurrent memory, provides a more effective implicit closure for this class of problems.

\bibliographystyle{jfm}
\bibliography{bibliography}

\end{document}